\documentclass[twocolumn,twocolappendix]{aastex701}
\usepackage{amsmath}
\usepackage{amssymb}
\usepackage{enumerate}
\usepackage[normalem]{ulem}
\usepackage{soul}
\usepackage{hyperref}

\newcommand{\be}{\begin{eqnarray}}
\newcommand{\ee}{\end{eqnarray}}

\newcommand{\msun}{\ensuremath{M_{\odot}}}

\newcommand{\lr}[1]{\left( #1 \right )}

\newcommand{\lp}[1]{\textbf{\textrm{\color{red} #1}}}
\renewcommand{\st}[1]{}
\renewcommand{\lp}[1]{#1}

\begin{document}

\title{Aspherical Remnants of Triple and Quadruple Detonations in Binary White Dwarfs}

%% The \author command is the same as before except it now takes an optional
%% argument which is the 16 digit ORCID. The syntax is:
%% \author[xxxx-xxxx-xxxx-xxxx]{Author Name}

\correspondingauthor{Logan Prust}
\email{lprust@flatironinstitute.org}

\author[0000-0002-7174-8273]{Logan J. Prust}
\affiliation{Kavli Institute for Theoretical Physics, University of California, Santa Barbara, CA, USA}
\affiliation{Center for Computational Astrophysics, Flatiron Institute, 162 Fifth Avenue, New York, NY 10010, USA}
\email{lprust@flatironinstitute.org}

\author[0000-0001-8038-6836]{Lars Bildsten}
\affiliation{Kavli Institute for Theoretical Physics, University of California, Santa Barbara, CA, USA}
\affiliation{Department of Physics, University of California, Santa Barbara, CA, USA}
\email{bildsten@kitp.ucsb.edu}

\author[0000-0002-1184-0692]{Samuel J. Boos}
\affiliation{Kavli Institute for Theoretical Physics, University of California, Santa Barbara, CA, USA}
\affiliation{Department of Physics \& Astronomy, The University of Alabama, Tuscaloosa, AL, USA}
\email{sjboos@crimson.ua.edu}

\begin{abstract}

White dwarfs which explode by the double-detonation mechanism may have a binary white dwarf donor which is subsequently ignited by its collision with the ejecta. This results in the destruction of the donor via either the triple- or quadruple-detonation mechanism, adding significant mass to the resulting ejecta as well as modifying its structure and composition. We simulate the evolution of supernova remnants resulting from such detonations in a variety of binary progenitors and compare them against a double detonation with a surviving donor. Because of the time delay between the detonations of the two white dwarfs, high-velocity ejecta from the first explosion governs the first few centuries of remnant evolution, whereas at later times the dense core resulting from the donor detonation drives both the forward and reverse shocks to larger radii. The collision between the highest-velocity ejecta of the primary explosion and the donor carves a conical wake into the ejecta, which persists into the remnant phase regardless of whether or not the donor detonates. 
\lp{Our suite of simulated remnants are found to exhibit multiple distinguishing features of the explosion properties: a distinct X-ray morphology in the thermal emission and iron lines for triple detonations and smaller remnants with centrally-concentrated emission for double detonations. The remnants are also varied in their elemental abundances and distributions, particularly for lighter elements, but these have limited observational utility and are sensitive to the properties of the progenitor binary.}
% 236 words

\end{abstract}


\section{Introduction} \label{sec:intro}

Type Ia supernovae (SNIa) may be triggered by the merger of two white dwarfs or by accretion onto a white dwarf (WD) from its binary donor \citep[for a review see][]{2019ApJ...887...68B,2025A&ARv..33....1R}. In the latter case, one proposed mechanism is the ``double detonation,'' which is triggered by the ignition of a shell of accreted helium. This sends a shock wave into the carbon/oxygen core, which converges at some point within the interior of the WD (though generally not at its center). At the convergence point, a second detonation is triggered, fusing the carbon-oxygen (CO) WD to radioactive $^{56}$Ni \citep{2014ApJ...785...61S}.

If the Roche lobe-filling donor is also a WD, the tightness of the binary corresponds to a high orbital velocity ($\gtrsim$800 km/s). For this reason, double detonations have been proposed as a formation channel for the population of hypervelocity white dwarfs observed by Gaia \citep{2018ApJ...865...15S} and SDSS \citep{2025arXiv250608081H}, one of which (D6-2) has been traced back to a supernova remnant \citep{2023OJAp....6E..28E}. Furthermore, hypervelocity WDs tend to occupy an unusual region of the Hertzsprung-Russell diagram consistent with heating due to interaction with supernova ejecta \citep{2019ApJ...887...68B}. To this end, there is a growing body of work on the effects of such an interaction on the surviving donor \citep{2015MNRAS.449..942P,2018ApJ...868...90T,2019ApJ...887...68B,2024ApJ...973...65W,2025arXiv250812529W}, with some finding that the interaction would cause the donor to detonate as well \citep{2015MNRAS.449..942P,2019ApJ...885..103T}. This is substantiated by the fact that several type Ia supernova remnants (SNRs) have been searched for surviving donors without success, such as Tycho \citep{2004Natur.431.1069R,2013ApJ...774...99K}, Kepler \citep{2018ApJ...862..124R}, SN 1006 \citep{2018MNRAS.479..192K,2022ApJ...933L..31S}, and SNR 0509-67.5 \citep{2023ApJ...950L..10S}. In such events, the ejecta mass is limited not by the Chandrasekhar mass, but by the total mass of the binary. This fact alone may be significant for studies of Ia remnants, as an assumed ejecta mass is often used to infer the local density of the interstellar medium (ISM). Indeed, it has been shown that the SNR 0509-67.5 is consistent with both Chandrasekhar-mass \citep{2022ApJ...938..121A} and sub-Chandrasekhar mass \citep{2019PhRvL.123d1101S,2025NatAs.tmp..135D} detonations. 

The nature of the donor detonation depends on the properties of the progenitor binary, falling into two categories. The collision between the ejecta may be strong enough to directly induce a catastrophic detonation wave when it first impacts the donor, resulting in a ``triple detonation'' scenario. If the donor is a helium WD, this is the likely way of igniting the donor \citep{2015MNRAS.449..942P, 2019ApJ...885..103T, 2024ApJ...972..200B}. On the other hand, if the donor is a CO WD, the shock wave may only be strong enough to ignite the donor's helium shell. In this case, the donor undergoes its own double detonation, so this is referred to as a ``quadruple detonation.'' Interestingly, \citet{2021MNRAS.503.4734P} showed that in some cases the detonation of the primary helium shell is sufficient to ignite the donor shell prior to central ignition of the primary.

Though the nuclear yields and ejecta masses of triple and quadruple detonations differ from double detonations, their spectra and peak brightnesses have been shown to be similar for donor masses $\lesssim 0.9$ $\msun$ \citep{2024ApJ...972..200B}. However, the remnant phase may offer a way to break the degeneracy between these different types of detonations. Here SNR 0509-67.5 is again of interest, as it has recently been shown to contain two distinct shells of calcium \citep{2025NatAs.tmp..135D}. This may indicate a double-detonation origin \citep{2025NatAs.tmp..135D}, though donor detonations have also been shown to produce at least two distinct calcium shells \citep{2024ApJ...972..200B}. Destruction of the donor would also create a shell of sulfur at low velocity, which has not been observed in SNR 0509-67.5, though the reverse shock may not have yet reached this location. A large fraction of the known SNIa remnants are slated for observations by the X-Ray Imaging and Spectroscopy Mission \citep{2024arXiv240814301X} in the coming years. This makes theoretical predictions of Ia SNR X-ray emission a priority, as the XRISM-Resolve spectrometer has already demonstrated the ability to probe the dynamics of core-collapse remnants such as SN1987A \citep{2025arXiv250507479X}, Cassiopeia A \citep{2025arXiv250504691V,2025arXiv250403268B,2025arXiv250323640S}, and N132D \citep{2025arXiv250403223G}.

Several groups have investigated the effect of interaction with a surviving donor on Ia remnants in the single-degenerate \citep{2012ApJ...745...75G,2016ApJ...833...62G} and double-degenerate \citep{2022ApJ...930...92F, 2025ApJ...982...60P, 2025arXiv251018800F} cases, showing that the interaction has a significant impact on SNR morphology. This may be responsible for the relatively degree of mirror symmetry exhibited by type Ia SNRs \citep{2011ApJ...732..114L}. In triple and quadruple detonations, much of the ejecta still interacts hydrodynamically with the donor prior to its destruction. The donor then detonates, effectively raising the ejecta mass and explosion energy as well as injecting a large amount of material with different nucleosynthetic yields than the primary.

It is important to know how the complex ejecta structure resulting from these processes translates to SNR morphology. To this end, we use 3D hydrodynamical simulations to calculate the evolution of remnants resulting from triple and quadruple detonations. Our initial conditions are provided by \citet[hereinafter BTS24]{2024ApJ...972..200B}, who simulated a suite of detonation models with varying WD masses and detonation mechanisms.

We organize this paper as follows. We discuss the state of the ejecta immediately following the detonation in section \ref{sec:thermo} and describe our numerical methods for evolving the fluid through the remnant phase in section \ref{sec:sproutsetup}. The dynamics of the resulting remnants are shown in section \ref{sec:dynamics}. In section \ref{sec:emission}, we present the composition of these objects and discuss the distinguishing features of each. We conclude and propose future directions in section \ref{sec:discussion}.

\section{Ejecta Thermodynamics} \label{sec:thermo}

\begin{figure*}
\begin{center}
\begin{tabular}{lll}
  \includegraphics[height=0.35\textwidth,trim={4.5cm 0.4cm 0.5cm 0},clip]{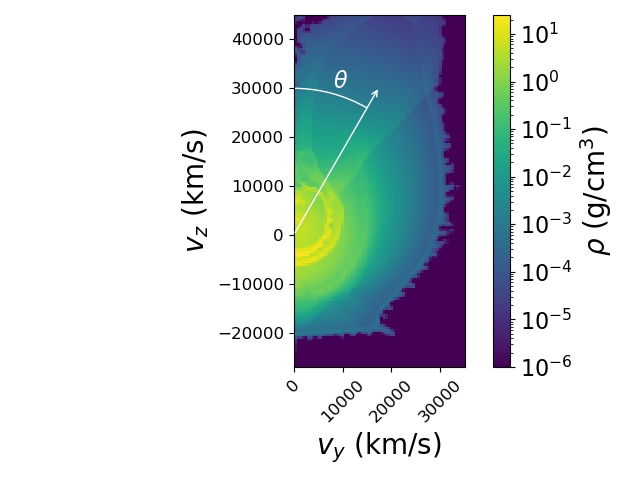} &
  \includegraphics[height=0.35\textwidth,trim={6.0cm 0.4cm 0.5cm 0},clip]{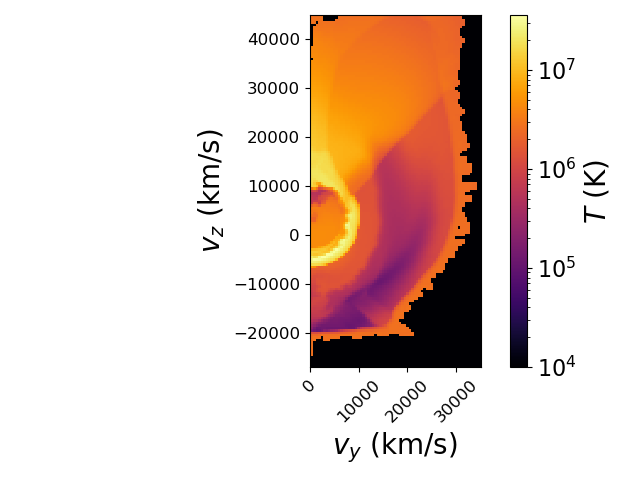} & \includegraphics[height=0.35\textwidth,trim={5.5cm 0.4cm 0.5cm 0.3cm},clip]{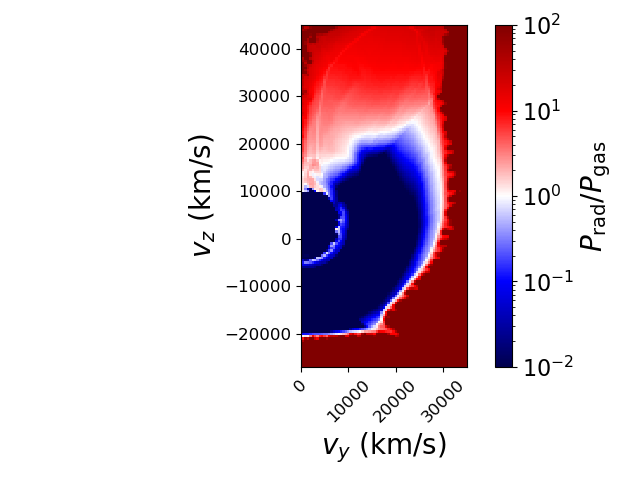}
\end{tabular}
\end{center}
\caption{Density \textit{(left)}, temperature \textit{(center)}, and ratio of radiation to gas pressure \textit{(right)} for the $1+0.7$ $\msun$ model at $t\approx 1$ minute after detonation. The bow shock and dense inner shell of ejecta are visible, as well as the shock-heated wake. We also show the definition of $\theta$, with $\theta=0$ along the center of the wake.}
\label{fig:detslices}
\end{figure*}

As the properties of the SNR are greatly influenced by the initial ejecta structure, we begin by describing the state of the ejecta about one minute after the detonation. We use the models described in BTS24, which are available in \citet{boos_2024_10515767}. These eight models include six in which the donor is detonated, one in which the donor survives, and one in which the exploding white dwarf does not have a companion (this scenario is used for the purpose of comparison). These models are listed in Table \ref{tab:detmodels} along with the total masses of several elements, which vary due to the differing nucleosynthetic yields. A central ignition did not occur in the $1+0.4$ $\msun$ model, as the donor is a He WD. Here the donor is directly ignited at its surface when it first collides with the ejecta. There are two $1.1+1$ $\msun$ models, one with a central ignition and one with a direct ignition, for the purposes of comparison. Note that ``direct ignition'' is synonymous with a triple detonation, and ``central ignition'' with a quadruple detonation. The simulations from BTS24 used WD progenitors that do not have He shells. (They show, however, that detonations of thin helium shells alter the observables arising from the the underlying core detonation to a very minor degree in the photospheric phase.) Thus, this work probes the nature of how the core ejecta evolves in the remnant phase, without consideration of the helium shell ejecta due to its low mass. However, \citet{2022ApJ...930...92F} showed that shell ashes can produce a protrusion in the forward and reverse shocks in a direction antipodal to the primary shell ignition point, lasting for a few centuries.

Several properties of the $1+0.7$ $\msun$ quadruple detonation model are shown in Fig.~\ref{fig:detslices}. Here several important features can be identified:
\begin{enumerate}
    \item The ejecta contains a dense shell of material at $v\approx10{,}000$ km/s owing to the detonation of the donor, with ejecta from the primary detonation at larger velocities.
    \item The fact that the primary was detonated off-center creates asymmetry in the ejecta regardless of the donor, with the lowest velocities (and highest densities) occurring along the $-z$-axis.
    \item Due to the time delay between the WD detonations, hydrodynamical interactions between the donor and ejecta have a substantial impact on the ejecta structure. Notably, the shock cone created by the donor can be seen as a diagonal line in the $\rho$ and $T$ plots in Fig.~\ref{fig:detslices}.
    \item The shock-heated ejecta contained within this ``wake'' is radiation pressure-dominated, in contrast to the vast majority of the ejecta \citep[as shown in][]{2025ApJ...982...60P,2025arXiv250719722K}.
    \item The ejecta within the wake is accelerated by its interaction with the donor. This effect is also seen in the surviving-donor models of both BTS24 and \citet{2025ApJ...982...60P}, but it is important to note that a wake forms in the high-velocity material regardless of the fate of the donor. The relation of this high-velocity component to observed high-velocity features remains of interest.
    \item Due to the 2D nature of the calculations performed in BTS24, the ignition point of the He shell on the primary was on the symmetry axis rather than at the point at which the accretion stream impacts the primary (as in 3D simulations such as \citet{2010ApJ...709L..64G} and \citet{2022MNRAS.517.5260P}). This means that the asymmetries created by the off-center detonation of the primary and by the presence of the donor are coaxial, which may not be the case in reality.
\end{enumerate}

\begin{figure}
    \centering
    \includegraphics[width=1.0\linewidth]{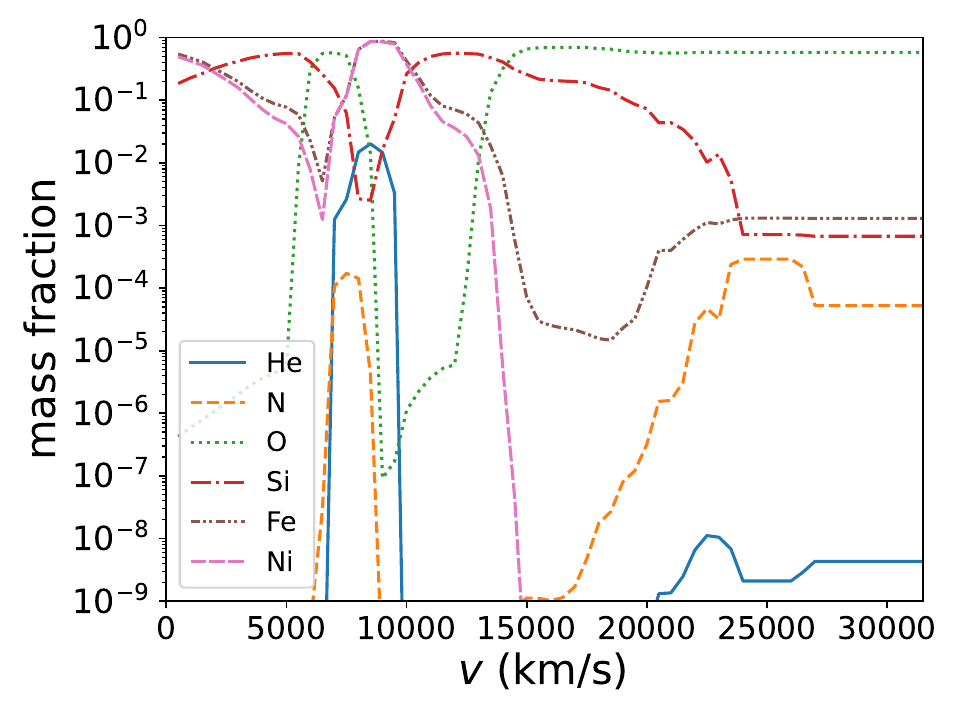}
    \caption{Radial profiles of the mass fractions $X_{i}$ along $\theta=90^{\circ}$ in the $1+0.7$ $\msun$ model at $t\approx 1$ min. Notably, significant $^{56}$Ni is found only at $v<15{,}000$ km/s. Ejecta from the donor detonation is primarily found at $v<10{,}000$ km/s.}
    \label{fig:1delements}
\end{figure}

In Fig.~\ref{fig:1delements}, we show the distribution of elements in the $1+0.7$ $\msun$ model along a radial ray at a right angle to the symmetry axis. Significant $^{56}$Ni is found only for $v<15{,}000$ km/s, with the shell of ejecta from the donor detonation at $5{,}000<v<10{,}000$ km/s. To study the thermodynamics of the ejecta moving forward, we consider the gas and radiation specific entropy
\be
s=\frac{k_B}{\mu m_p}\ln \lr{\frac{T^{3/2}}{\rho}}+\frac{4a_{r}T^3}{3\rho}. \label{eq:entropy}
\ee
The change in entropy due to the decay of $^{56}$Ni to $^{56}$Co, and subsequently to $^{56}$Fe, is given by
\be
T\frac{ds}{dt}=\frac{X_{\rm Ni} \epsilon_{\rm Ni}}{56 m_p t_{\rm Ni}} e^{-t/t_{\rm Ni}} + \frac{X_{\rm Co} \epsilon_{\rm Co}}{56 m_p t_{\rm Co}} e^{-t/t_{\rm Co}}. \label{eq:dsdt}
\ee
Here the $X_{i}$ are the mass fractions, $\epsilon_{\rm Ni}=1.72$ MeV and $\epsilon_{\rm Co}=3.49$ MeV are energies released in each decay \citep{1994ApJS...92..527N}, $t_{\rm Ni}=8.77$ days and $t_{\rm Co}=111$ days are the $e$-folding times, $T$ is the temperature of both the gas and radiation, $\rho$ is the gas density, $a_{r}$ is the radiation constant, $k_{B}$ is the Boltzmann constant, $m_{p}$ is the proton mass, and $\mu=2$ is the mean molar mass. We numerically integrate (\ref{eq:entropy}) and (\ref{eq:dsdt}) forward in time, neglecting radiative diffusion. The temperature is determined at each step by performing a root find on (\ref{eq:entropy}), and the density evolves simply as $\rho\propto t^{-3}$. This yields the gas pressure $P_{\rm gas}=\rho k_{B}T/\mu m_{p}$ and radiation pressure $P_{\rm rad}=a_{r}T^{4}/3$.

\begin{figure}
    \centering
    \includegraphics[width=1.0\linewidth]{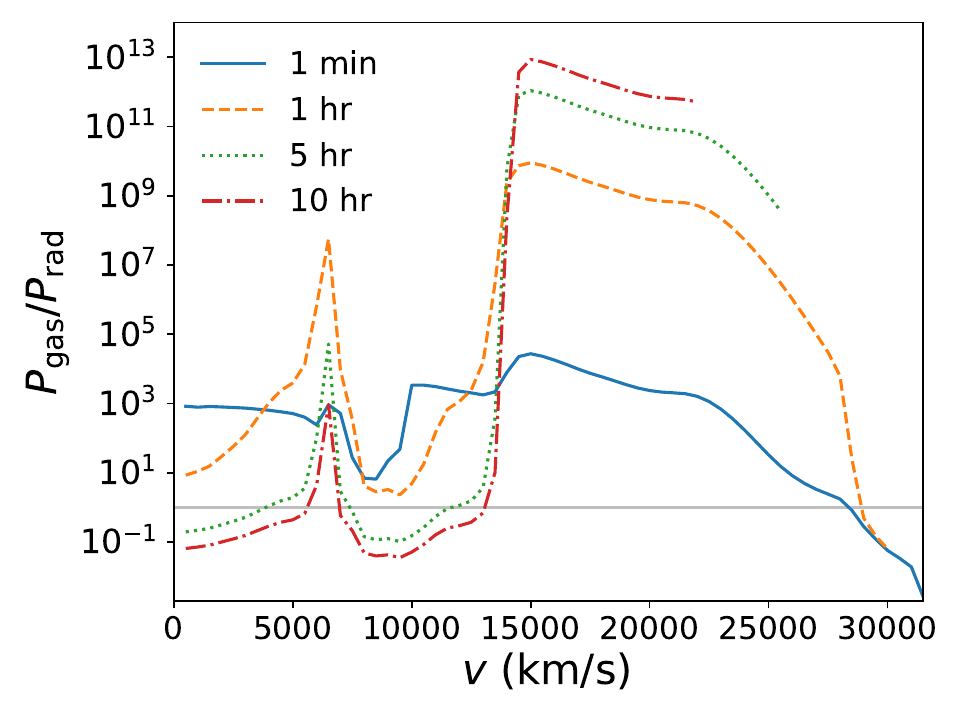}
    \includegraphics[width=1.0\linewidth]{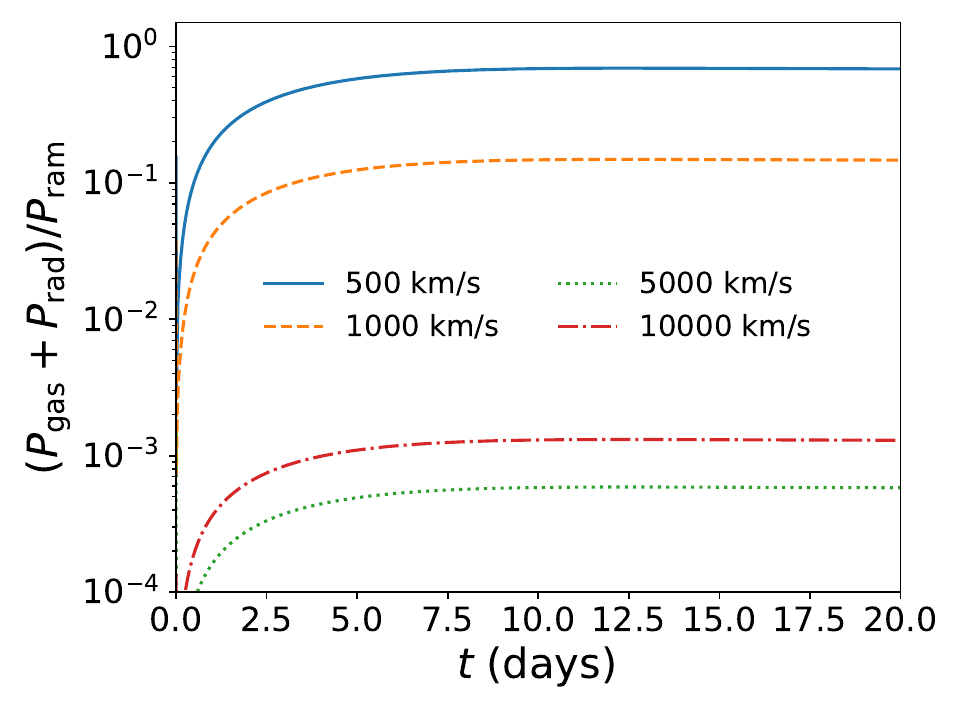}
    \caption{\textit{(Top)} Ratio of gas pressure to radiation pressure along $\theta=90^{\circ}$ in the $1+0.7$ $\msun$ model, showing only the optically-thick ejecta ($\tau>c/v$) assuming complete ionization. The $^{56}$Ni-rich ejecta quickly heats to become radiation-pressure dominated. \textit{(Bottom)} A measure of the homology of fluid parcels at various velocities. All ejecta with $v>1{,}000$ km/s maintains a ram pressure which exceeds the gas and radiation pressure by an order of magnitude.}
    \label{fig:1dhomology}
\end{figure}

The resulting evolution of the ratio $P_{\rm rad}/P_{\rm gas}$ is shown in the top panel of Fig.~\ref{fig:1dhomology}, where we show only optically-thick ejecta (with optical depth $\tau>c/v$) assuming complete ionization. While the $^{56}$Ni-rich ejecta generally becomes radiation pressure-dominated after a few hours, the rest is quickly cooled by adiabatic expansion and becomes highly gas pressure-dominated. 
%\citet{2025arXiv250719722K} showed that all of the ejecta will become radiation pressure-dominated due to $^{56}$Ni heating after several hours. 
The outermost ejecta was the recipient of a large amount of energy from the detonation wave, and remains dominated by radiation pressure (at least until it becomes optically thin).

We are also equipped to determine the degree to which the fluid can be approximated as homologous by comparing the gas, radiation, and ram pressure. In the bottom panel of Fig.~\ref{fig:1dhomology}, we show the ratio $(P_{\rm gas}+P_{\rm rad})/P_{\rm ram}$ for several fluid parcels. As we neglect radiative transfer, this is actually an upper limit on the ratio. We see that, with the exception of very low velocities ($v<1{,}000$ km/s), the gas can be well-approximated as homologous. This is a small fraction of the fluid and has little effect on the remnant evolution regardless, so we conclude that the detonation models can be safely approximated as homologous until the ISM becomes dynamically significant.

\begin{table*}
%\begin{center}
\caption{Summary of the detonation models from BTS24, including the primary and donor masses $M_{1}$ and $M_{2}$ as well as the total masses of several notable elements. All masses listed are in units of $\msun$. We combine the masses of Fe and $^{56}$Ni as all $^{56}$Ni will decay to iron by the remnant phase.}
\begin{tabular}{cc|c|ccccccc}
\hline
\hfill $M_{1}$ & $M_{2}$ & Notes & $M_{\rm He}$ & $M_{\rm N}$ & $M_{\rm O}$ & $M_{\rm S}$ & $M_{\rm Si}$ & $M_{\rm Fe}+M_{\rm ^{56}Ni}$ \\
\hline
\hfill 1.00 & ---  & isolated WD & $9.79\times 10^{-3}$ & $1.23\times 10^{-6}$ & 0.079 & 0.097 & 0.175 & 0.60 \\
\hfill 1.00 & 0.40 & He donor & $1.68\times 10^{-1}$ & $2.35\times 10^{-6}$ & 0.077 & 0.100 & 0.177 & 0.75 \\
\hfill 1.00 & 0.70 & central ignition & $9.66\times 10^{-3}$ & $2.80\times 10^{-5}$ & 0.343 & 0.205 & 0.411 & 0.63 \\
\hfill 1.00 & 0.70 & surviving donor & $9.21\times 10^{-3}$ & $1.34\times 10^{-6}$ & 0.079 & 0.096 & 0.173 & 0.56 \\
\hfill 1.00 & 0.90 & central ignition & $1.55\times 10^{-2}$ & $2.61\times 10^{-6}$ & 0.177 & 0.213 & 0.388 & 1.02 \\
\hfill 0.85 & 0.80 & central ignition & $2.07\times 10^{-3}$ & $1.55\times 10^{-5}$ & 0.360 & 0.270 & 0.529 & 0.39 \\
\hfill 1.10 & 1.00 & central ignition & $2.60\times 10^{-2}$ & $9.14\times 10^{-7}$ & 0.089 & 0.142 & 0.248 & 1.50 \\
\hfill 1.10 & 1.00 & direct ignition & $2.51\times 10^{-2}$ & $6.06\times 10^{-7}$ & 0.103 & 0.157 & 0.284 & 1.44 \\
\hline
\label{tab:detmodels}
\end{tabular}
%\end{center}
\end{table*}

% \begin{table*}
% \begin{center}
% \caption{Relevant velocities and timescales at the time of detonation. Listed are the relative velocities of the WDs $v_{\rm rel}$, rest-frame velocity of the companion $v_2$, delay between core detonations $t_{\rm delay}$, and orbital period $T_{\rm orb}$.}
% \begin{tabular}{ccc|ccccc}
% \hline
% \hfill $M_{1}$ ($\msun$) & $M_{2}$ ($\msun$) & $a$ ($10^3$ km) & $v_{\rm rel}$ (km/s) & $v_{2}$ (km/s) & $t_{\rm delay}$ (s) & $T_{\rm orb}$ (s) & $t_{\rm delay}/T_{\rm orb}$ \\
% \hline
% \hfill 1.00 & ---  & ---  & ---  & ---  & ---  & ---   & ---   \\
% \hfill 1.00 & 0.40 & 38.2 & 2212 & 1580 & 2.40 & 108.5 & 0.022 \\
% \hfill 1.00 & 0.70 & 21.8 & 3226 & 1898 & 3.35 & 46.8  & 0.072 \\
% \hfill 1.00 & 0.70 & 21.8 & 3226 & 1898 & ---  & ---   & ---   \\
% \hfill 1.00 & 0.90 & 18.1 & 3743 & 1970 & 2.90 & 35.4  & 0.082 \\
% \hfill 0.85 & 0.80 & 20.8 & 3254 & 1676 & 3.65 & 43.6  & 0.084 \\
% \hfill 1.10 & 1.00 & 15.7 & 4225 & 2213 & 2.90 & 28.6  & 0.101 \\
% \hfill 1.10 & 1.00 & 15.6 & 4239 & 2220 & 1.10 & 28.3  & 0.039 \\
% \hline
% \label{tab:velocities}
% \end{tabular}
% \end{center}
% \end{table*}

\section{Numerical Setup for Remnant Evolution} \label{sec:sproutsetup}

We perform 3D hydrodynamical simulations of the remnant evolution using the moving-mesh code Sprout \citep{sprout}.
%, which solves the Euler equations
% \be
% \ddt{\rho} + \grad\cdot\rho\vel &=& 0, \label{eq:continuity}\\
% \ddt{\rho\vel} + \grad\cdot\rho\vel\vel^{T} + \grad P &=& 0, \label{eq:momentum}\\
% \ddt{u} + \grad\cdot\left(u + \rho v^{2}/2 + P\right)\vel &=& 0 \label{eq:energy}
% \ee
% for gas with fluid velocity $\vel$ and internal energy $u$. 
We use a numerical setup identical to that of \citet{2025ApJ...982...60P}, which we summarize below, with the exception that we use third-order Runge-Kutta time integration.
We model a quadrant of the ejecta, i.e. $\theta\in [0,\pi]$ and $\phi\in [0,\pi/2]$, as this is sufficient to capture the relevant 3D effects. Our Cartesian grid contains $N_{x}\times N_{y}\times N_{z}=512\times 512\times 1024=268$ million cells, with the $\theta=0$ pole aligned with the $+z$-axis and reflective boundary conditions on the $x=0$ and $y=0$ planes.
Sprout uses a Cartesian grid which expands at the same rate as the SNR, mitigating numerical diffusion. Carbuncle instabilities are encountered along the $\theta=180^{\circ}$ pole where the grid aligns with the forward shock \citep{1998JCoPh.145..511S}, though these have a small effect on the remnant morphology.

The initial size of our domain is roughly twice the linear size of the ejecta, with the center point for mesh expansion coinciding with the location of the supernova. The ISM density is set to $\rho_{\rm ISM}=6.31\times 10^{-25}$ g/cm$^{3}$, similar to that inferred by \citet{2022ApJ...938..121A} for the SNIa remnant SNR 0509-67.5. We begin our simulations at $t=10$ yrs, when the density of the outer layers of the ejecta is comparable to that of the ISM, and continue to $t=3000$ yrs. \lp{The ejecta is assumed to be homologous prior to $t=10$ yrs, as shown in the previous section, so its density as a function of velocity can be directly mapped from the detonation models of BTS24 into Sprout. The internal energy is not mapped from the detonation models; rather,} the gas pressure is chosen to be $P=10^{-5}\rho~(50{,}000~{\rm km/s})^{2}$ so that it is negligible compared to the ram pressure for both the ejecta and ISM \lp{(until the gas is later shocked, resetting its internal energy)}. The gas is treated as ideal with specific heat ratio $\gamma=5/3$. \lp{We perform pure hydrodynamical calculations, neglecting self-gravity, cosmic rays, and magnetic fields. We also assume that radiative losses are not yet dynamically significant, which occurs at roughly}
\begin{equation}
    t_{\rm rad} \approx (3.61\times10^4~{\rm yr})E_{51}^{3/14}/e\zeta_{m}^{5/14}n_{0}^{4/7}
\end{equation}
\lp{\citep{1988ApJ...334..252C}. Here $E_{51}$ is the explosion energy in units of $10^{51}$ ergs, $\zeta_{m}$ is a factor which depends on metallicity, and $n_{0}$ is the ISM number density. For our remnants, we estimate $t_{\rm rad}\approx 20{,}000$ yrs, which is much longer than our integration time of 3,000 yrs.} We track the mass fractions of all elements listed in Table \ref{tab:detmodels} as passive tracers. The fork of Sprout used for these calculations is publicly available \citep{soham_mandal_2025_15595613}.

\section{Remnant Dynamics} \label{sec:dynamics}

\begin{figure*}[p!]
    %\centering
    \hspace{1.1cm}\includegraphics[width=0.83\linewidth,trim={0 0 0 0},clip]{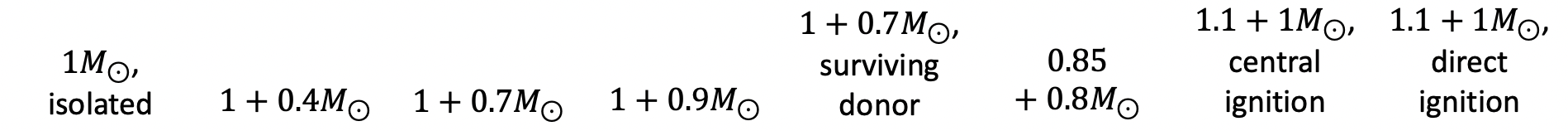}
    \includegraphics[width=1.0\linewidth,trim={1.9cm 1.65cm 0.2cm 1.8cm},clip]{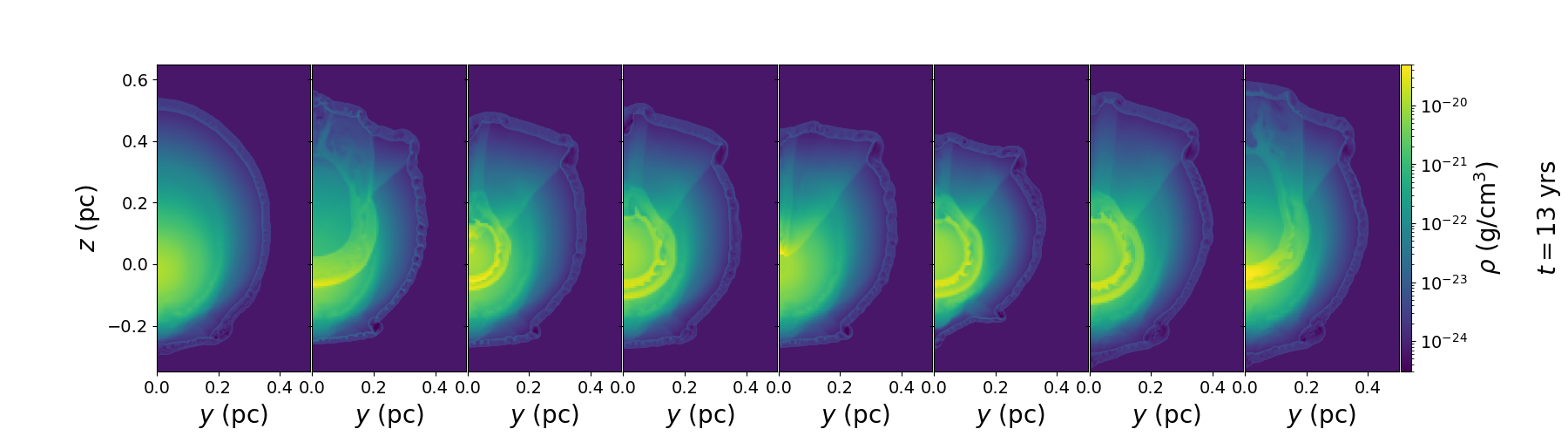}
    \includegraphics[width=1.0\linewidth,trim={1.9cm 1.65cm 0.2cm 1.7cm},clip]{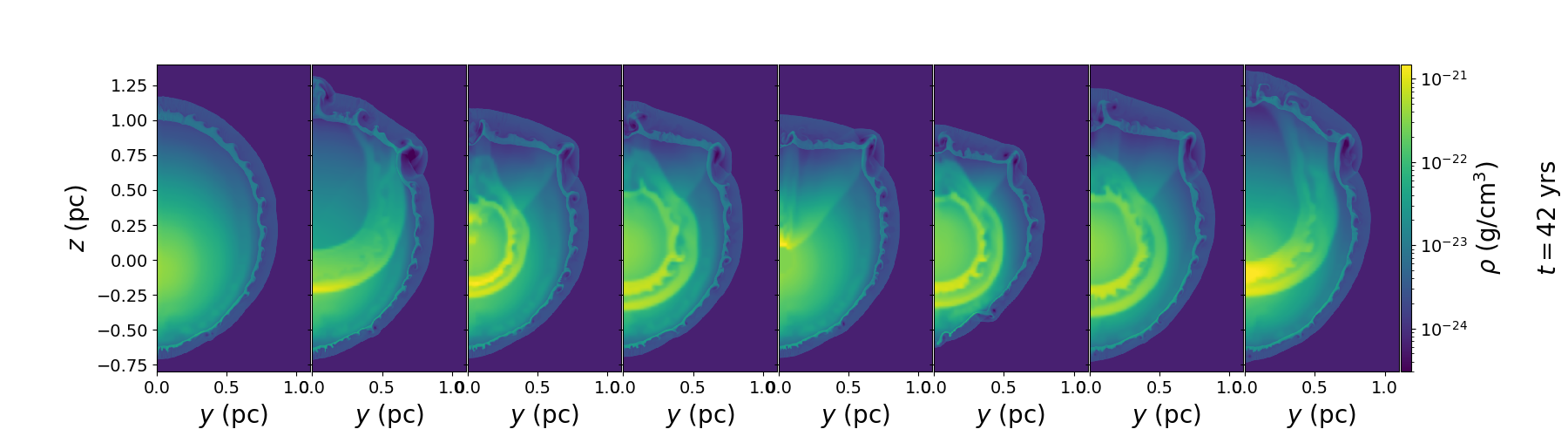}
    \includegraphics[width=1.0\linewidth,trim={1.9cm 1.65cm 0.2cm 1.7cm},clip]{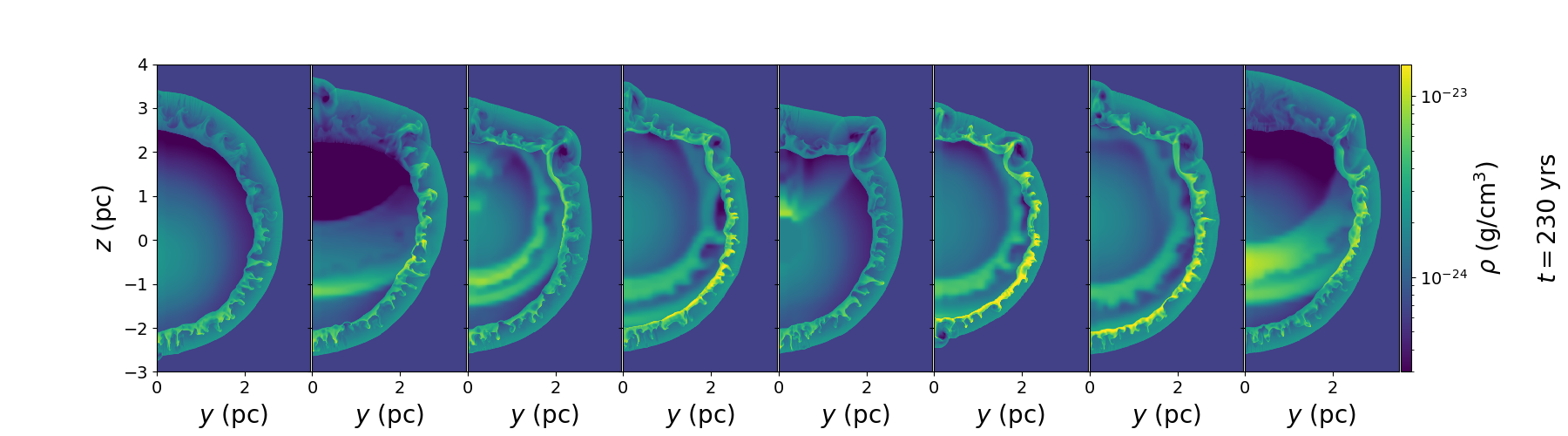}
    \includegraphics[width=1.0\linewidth,trim={1.9cm 1.65cm 0.2cm 1.7cm},clip]{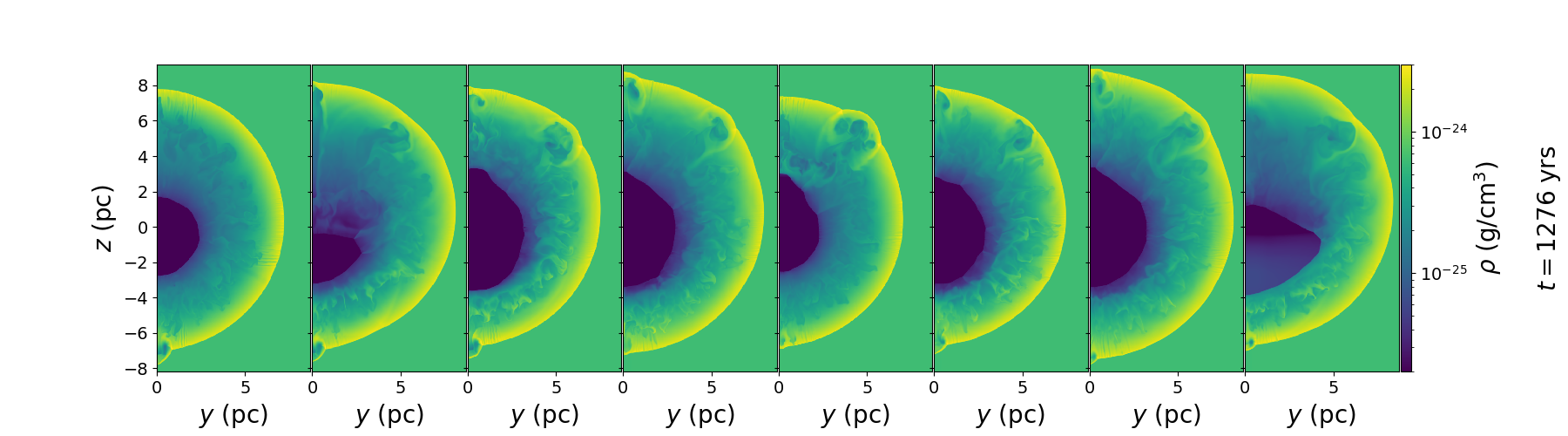}
    \includegraphics[width=1.0\linewidth,trim={1.9cm 1.65cm 0.2cm 1.7cm},clip]{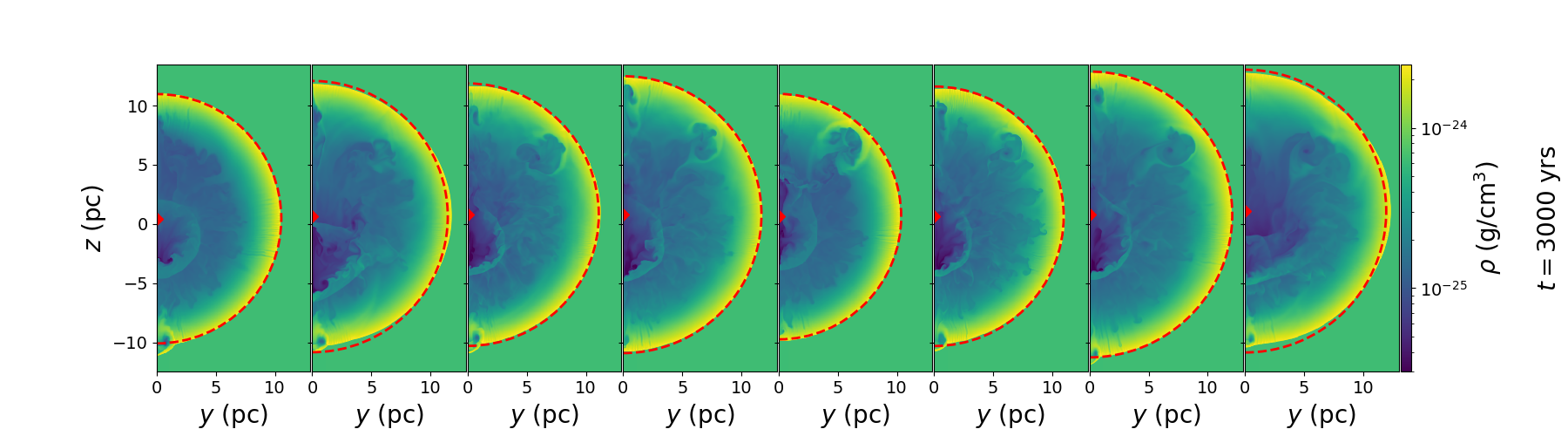}
    \caption{Density slices for each SNR on the $x=0$ plane at $t=13$, 42, 230, 1276, and 3000 yrs. The red dashed circles in the bottom row represent the spherical fits discussed in section \ref{sec:forwardshock}, with their centers marked by red diamonds.}
    \label{fig:rhogrid}
\end{figure*}

In Fig.~\ref{fig:rhogrid} we show density slices of each of the eight detonation models (columns) at five epochs (rows): $t=13$, 42, 230, 1276, and 3000 yrs. We can immediately identify several important features of these SNRs. The high-velocity ejecta along the $+z$-axis owing to hydrodynamical interactions between the primary ejecta and donor creates initial protrusions in the forward shock (FS) and reverse shock (RS) which last at least a few hundred years, regardless of whether or not the donor detonated. This is particularly pronounced at the edge of the wake, where large Rayleigh-Taylor plumes create structures that last for thousands of years. The isolated WD remnant is understandably more spherical, though the ejecta velocity remains angle-dependent (varying by a factor of $\approx$2 at early times) due to the off-center ignition of the CO core. Most models exhibit a carbuncle instability along the $-z$-axis to some degree.

All triple- and quadruple-detonation models exhibit an inner shell of dense ejecta. As the FS slows, this shell is swept up by the reverse shock after a few centuries. However, the shell has far less radial symmetry in the triple detonation models. Here the shorter time delay between the detonations did not allow for much of the primary ejecta to pass by the donor prior to the donor detonation. After $\approx$1000 years, the varying ejecta structures of these SNRs leads to a diversity of RS shapes (fourth row of Fig.~\ref{fig:rhogrid}), which then converge to create bounce shocks (fifth row). In the remainder of this section, we take a quantitative look at the features discussed above.

\begin{figure}
    \centering
    \includegraphics[width=1.0\linewidth]{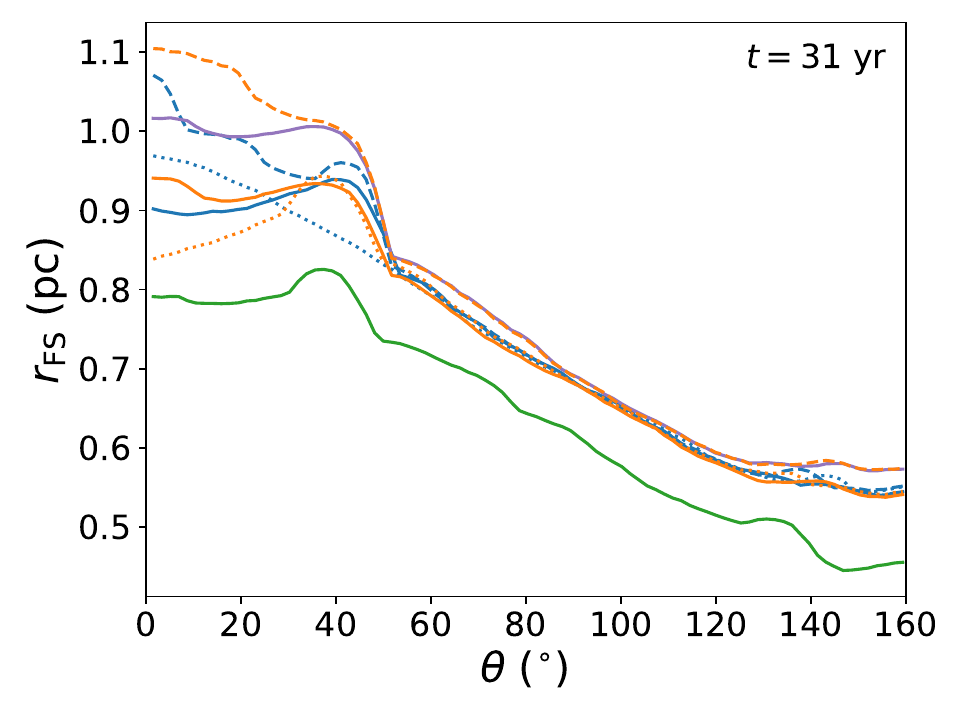}
    \includegraphics[width=1.0\linewidth]{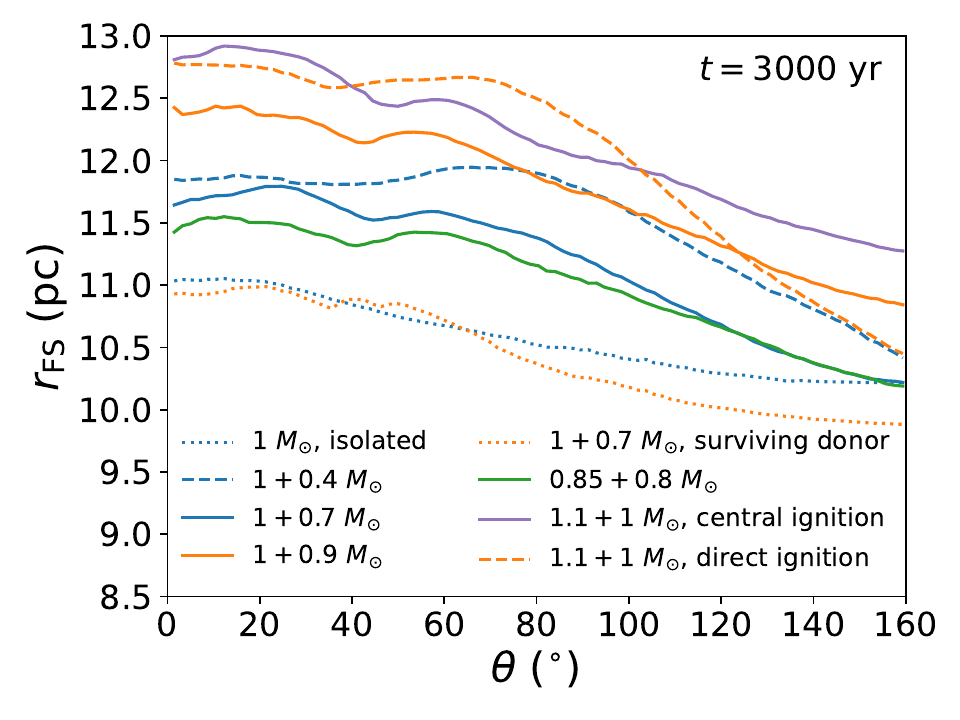}
    \includegraphics[width=1.0\linewidth]{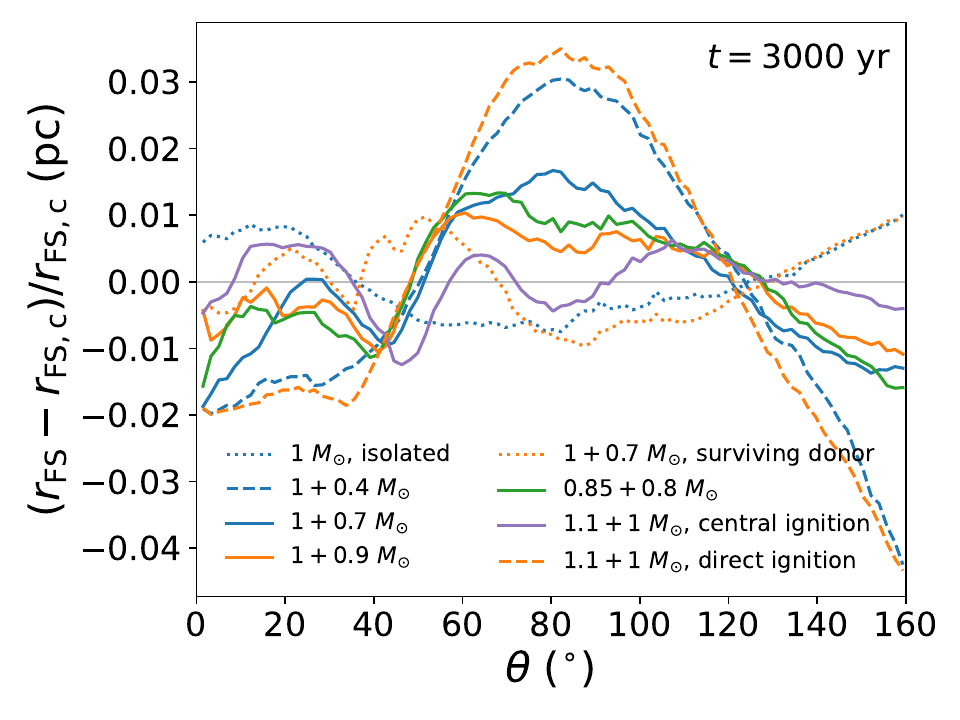}
    \caption{Forward shock radius vs angle at $t=31$ yr \textit{(top)} and $t=3000$ yr \textit{(center)} for $\rho_{\rm ISM}=6.31\times 10^{-25}$ g/cm$^{3}$. At early times, $r_{\rm FS}$ scales with the primary mass (outside of the wake), while at late times there is a bifurcation between the double detonations and the rest. \textit{(Bottom)} Fractional variation of the FS radius from the spherical fit, measured from the center of the spherical fit $c_{\rm FS,s}$. The variation is most pronounced for the triple detonations, particularly at the equatorial plane. The legend is the same for all panels.}
    \label{fig:FSmult}
\end{figure}

\begin{figure}
    \centering
    \includegraphics[width=1.0\linewidth]{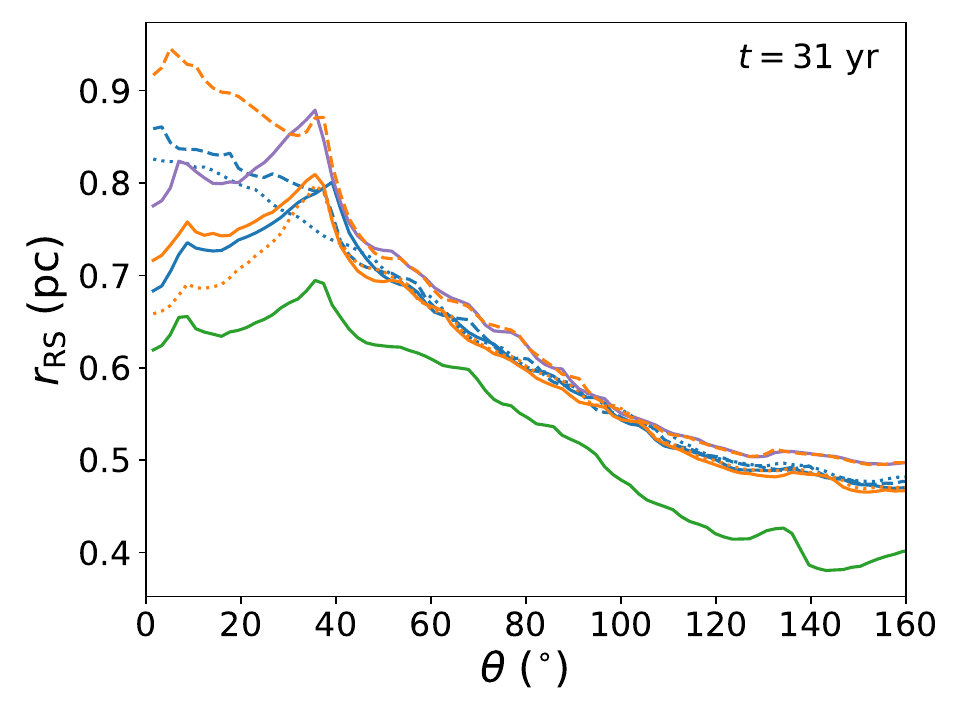}
    \includegraphics[width=1.0\linewidth]{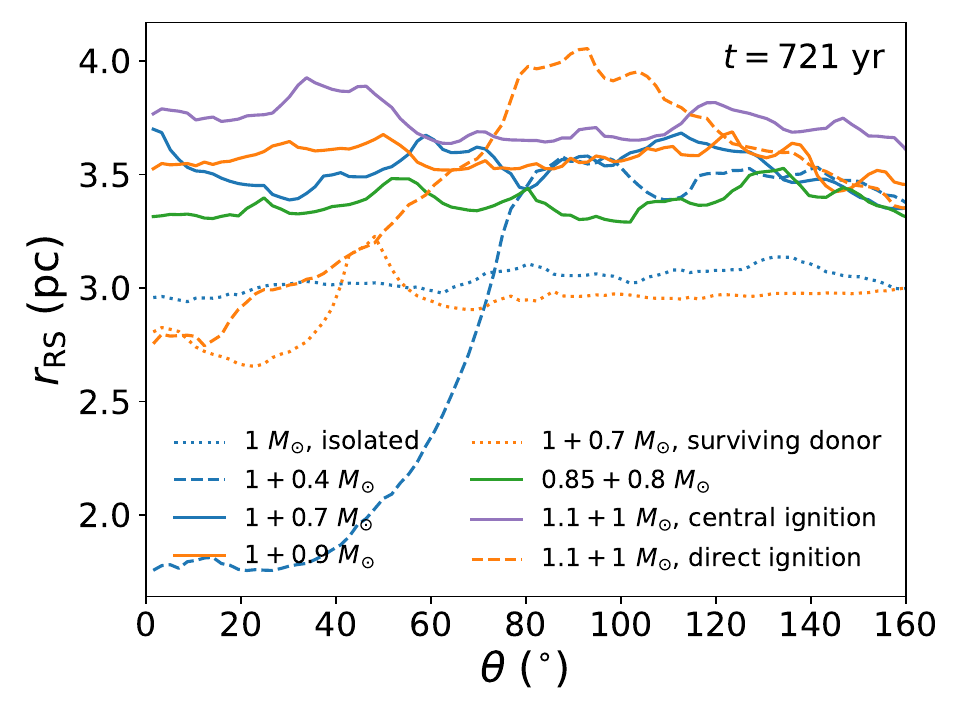}
    \caption{Reverse shock radius vs angle at $t=31$ yr \textit{(top)} and $t=721$ yr \textit{(bottom)}. As with the forward shock, $r_{\rm RS}$ scales with the primary mass at early times and the donor mass at late times. The RS penetrates the wake more easily in the surviving donor and direct detonation cases.}
    \label{fig:RSmult}
\end{figure}

\begin{figure}
    \centering
    \includegraphics[width=1.0\linewidth]{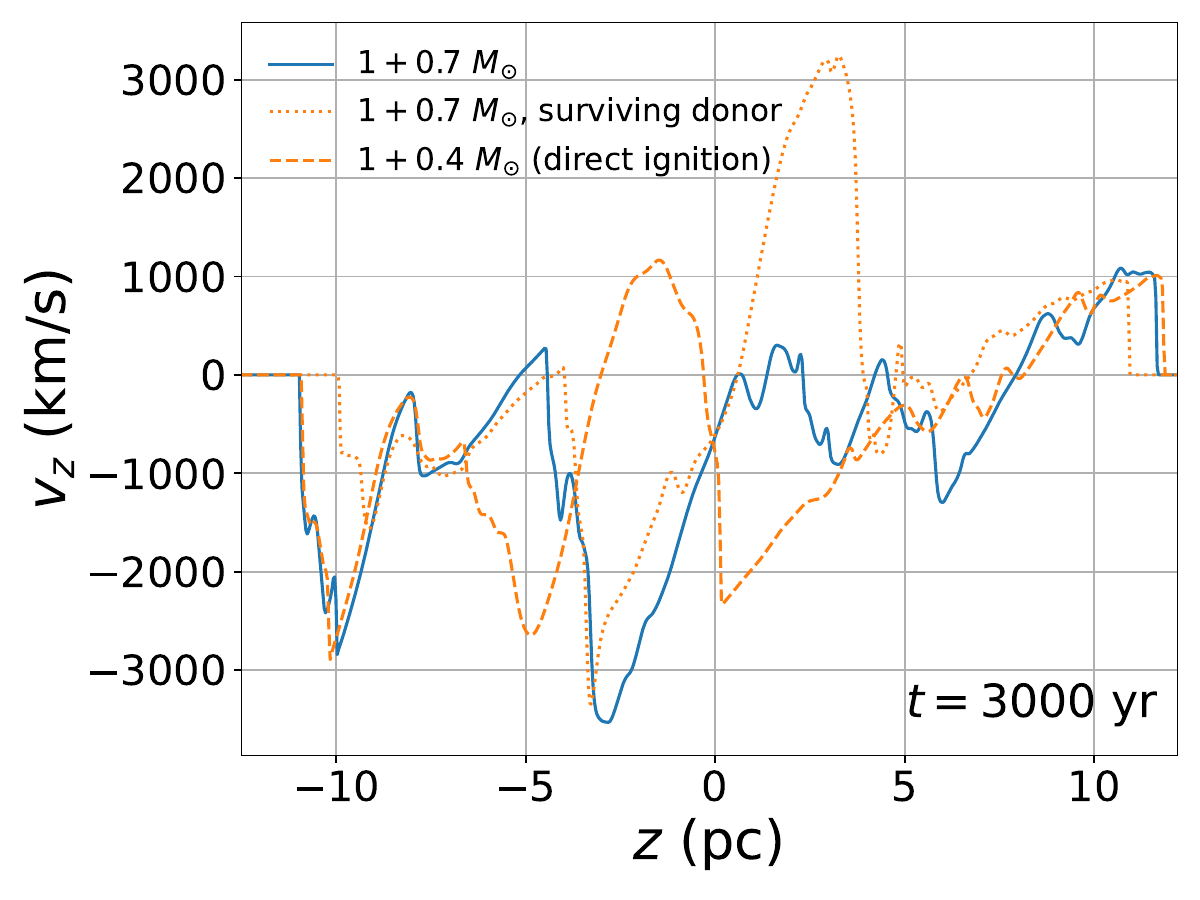}
    \caption{The post-bounce velocity profiles within the core differ significantly between the surviving donor (purple), triple detonation (orange), and quadruple detonation (green) cases.}
    \label{fig:velcompare}
\end{figure}

\begin{table*}
\begin{center}
\caption{Parameters of the spherical fit to the shape of each remnant (discussed in section \ref{sec:forwardshock}) at $t=3000$ yrs, including its radius $r_{\rm FS,s}$ and the offset between its center and the point of the supernova $c_{\rm FS,s}$.}
\begin{tabular}{cc|c|cc}
\hline
\hfill $M_{1}$ ($\msun$) & $M_{2}$ ($\msun$) & Donor Detonation Mechanism & $r_{\rm FS,s}$ (pc) & $c_{\rm FS,s}$ (pc) \\ % & $R^{2}$ \\
\hline
\hfill 1.00 & --- & none (isolated WD)       & 10.53 & 0.44 \\ % & 0.96 \\
\hfill 1.00 & 0.40 & direct                 & 11.46 & 0.62 \\ % & 0.77 \\
\hfill 1.00 & 0.70 & central                & 11.09 & 0.78 \\ % & 0.95 \\
\hfill 1.00 & 0.70 & none (surviving donor) & 10.36 & 0.61 \\ % & 0.98 \\
\hfill 1.00 & 0.90 & central                & 11.70 & 0.78 \\ % & 0.98 \\
\hfill 0.85 & 0.80 & central                & 10.96 & 0.65 \\ % & 0.95 \\
\hfill 1.10 & 1.00 & central                & 12.07 & 0.80 \\ % & 0.99 \\
\hfill 1.10 & 1.00 & direct                 & 11.95 & 1.08 \\ % & 0.88 \\
\hline
\label{tab:spherical_fits}
\end{tabular}
\end{center}
\end{table*}

\subsection{Forward Shock} \label{sec:forwardshock}

Fig.~\ref{fig:FSmult} shows the FS radius versus polar angle at $t=31$ yrs (top panel) and $t=3000$ yrs (center panel), neglecting $\theta>160^{\circ}$ due to carbuncle instabilities. The legend is the same for both panels and \lp{uses a different line style for each type of detonation.} At early times, the wake due to interaction with the donor has a substantial effect on the FS, causing a protrusion at $\theta\approx 40^{\circ}$ \citep[consistent with][]{2025ApJ...982...60P}. Within the wake, there is no clear ordering of the FS radius $r_{\rm FS}$ by detonation mechanism, however outside of the wake there is a clear trend. Those with a 1 $\msun$ primary are approximately equal, with the 1.1 $\msun$ primaries just above and the 0.85 $\msun$ primary just below, demonstrating that at early times $r_{\rm FS}$ scales with the primary mass. At late times, this trend is broken as the dynamics are now influenced by initially low-velocity material from the donor detonation. Here the models are bifurcated: SNRs resulting from the destruction of only one WD are significantly smaller than the rest. This is what one would intuitively expect, as their total ejecta masses are lower.

The off-center detonation of the primary is clearly demonstrated in both panels, with $r_{\rm FS}$ varying by $\approx$50\% at early times even for the isolated WD detonation. This effect is retained at late times to varying degrees for all detonation models. Here the SNRs have become more spherical, with the effects of the wake mostly washed out. The FS can be fit to a sphere, parameterized by a radius $r_{\rm FS,s}$ and the offset between the center of the sphere and the point of the supernova $c_{\rm FS,s}$. The fit parameters at $t=3000$ yrs are listed in Table \ref{tab:spherical_fits} and are overplotted in Fig.~\ref{fig:rhogrid}. Here we see quantitatively that $r_{\rm FS,s}$ is lower for the double detonations and that the geometric centers of the 3000-year-old remnants are offset from the supernovae by $\approx$ 0.5--1 pc. However, the spherical approximation is less valid for the direct detonations which tend to be wider along the equatorial plane by a few percent, as shown in the bottom panel of Fig.~\ref{fig:FSmult}.

\subsection{Reverse Shock} \label{sec:reverseshock}

\lp{The gas entropy is used to determine which gas parcels have been shocked: we define shocked gas as that which has pseudoentropy at least 4 times larger than the unshocked ISM:}
\begin{equation}
    \left.\frac{P}{\rho^{\gamma}}\right|_{\rm shocked} \geq 4 \left.\frac{P}{\rho^{\gamma}}\right|_{\rm ISM}.
\end{equation}
\lp{By inspection, all shocked gas has pseudoentropy exceeding than this value. The inner and outer boundaries of the shocked gas then define the reverse and forward shocks, respectively.} At early times, the RS shape traces that of the FS, thus exhibiting the same scaling with primary mass as well as the other features discussed above. Fig.~\ref{fig:RSmult} shows the reverse shock radius $r_{\rm RS}$ versus polar angle at two epochs; we choose earlier epochs for the RS as by $t=3000$ yrs it has long since reached the center of the remnant. We see that at $t=721$ yrs---which is after the RS has swept up the material from the donor detonation---the same bifurcation occurs as in the FS, with $r_{\rm RS}$ roughly half a parsec larger for the triple- and quadruple-detonation remnants than for double detonations. However, unlike the FS, the RS has ``forgotten'' the angle dependence of the high-velocity ejecta, and has become roughly spherical (with some notable exceptions) and centered on the location of the SN. 

As seen in Fig.~\ref{fig:rhogrid}, the structure of the wake can vary markedly between detonation mechanisms, and this is significant in the dynamics of the RS. For the surviving donor model, which does not contain an inner shell of dense ejecta but rather a single clump of material stripped from the donor, the RS traverses quickly within the wake. This is consistent with previous work on Ia SNRs following interaction with a donor \citep{2022ApJ...930...92F,2025ApJ...982...60P}. A similar phenomenon is seen in the triple detonation models, as here the donor ejecta shell is far more asymmetrical and nearly nonexistent at small $\theta$. This allows the RS to travel through the wake with little impedance. The effect is far more pronounced for the exploding 0.4 $\msun$ helium WD donor, which is likely because it simply provides less mass to impede the RS than the 1 $\msun$ CO donor.

The core of the remnant following the convergence of the RS at its center (bottom row of Fig.~\ref{fig:rhogrid}) also differs morphologically based on detonation mechanism, with the bounce shocks displaying qualitatively different shapes. Fig.~\ref{fig:velcompare} shows the velocity profile along the $z$-axis at $t=3000$ yrs for several remnants. We show one double, one triple, and one quadruple detonation, as each of these is representative of its type of explosion. The double detonation results in a plume of material being driven at high velocity in the $+z$-direction into the wake, reproducing a conclusion of \citet{2025ApJ...982...60P}. Quadruple detonations lack this feature but otherwise closely match the profile of the double detonations. Triple detonations are unique in that their profiles appear to be shifted to lower $z$ (both in Fig.~\ref{fig:velcompare} and in the bottom row of Fig.~\ref{fig:rhogrid}), likely owing to their very low densities at $z>0$ at early times. As discussed above, the RS passes through the wake and crosses the origin quickly in these models, shifting the location of the RS convergence point away from that of the original explosion.

\section{X-Ray Tomography} \label{sec:emission}

X-rays provide an excellent tool to study the dynamics of SNRs, as remnants are optically thin to X-rays. For a review of SNR X-ray radiation mechanisms, see \citet{2012A&ARv..20...49V}. The emission can be roughly separated into line emission and continuum thermal bremsstrahlung. 
%Though SNRs are also sources of radio synchrotron emission, X-ray synchrotron radiation is generally significant only in the presence of a pulsar wind nebula.

\subsection{Continuum Thermal Emission} \label{sec:continuum}

\begin{figure*}
    \centering
    \includegraphics[width=0.81\linewidth,trim={0.9cm 0 0 0},clip]{figures/labels3.png}
    \includegraphics[width=1.0\linewidth,trim={1cm 2.6cm 0.5cm 2.5cm},clip]{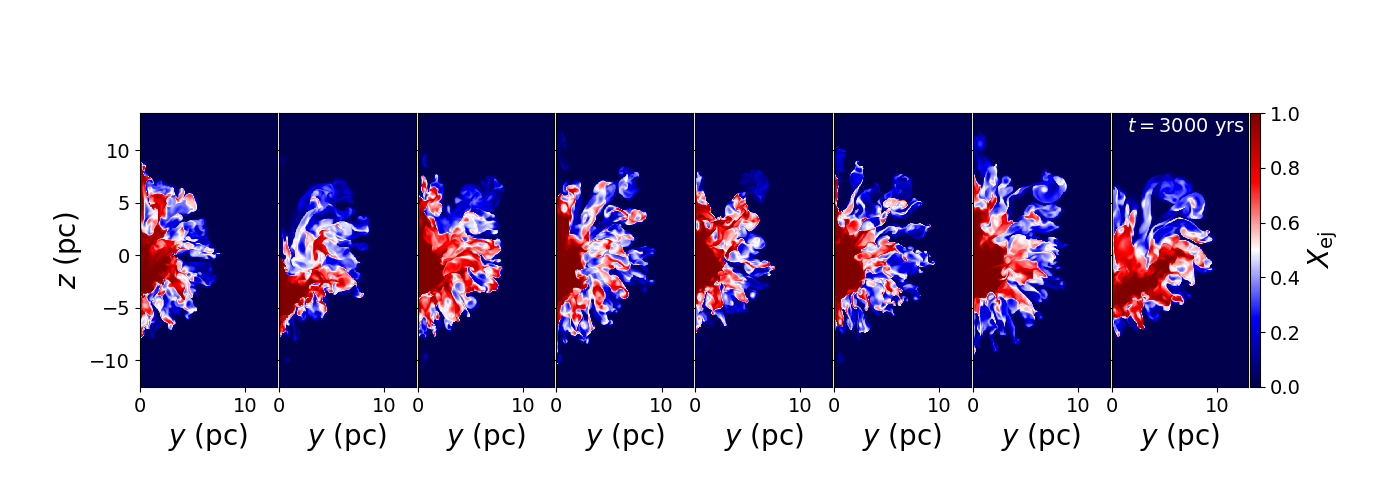}
    \includegraphics[width=1.0\linewidth,trim={1cm 2.6cm 0.5cm 2.8cm},clip]{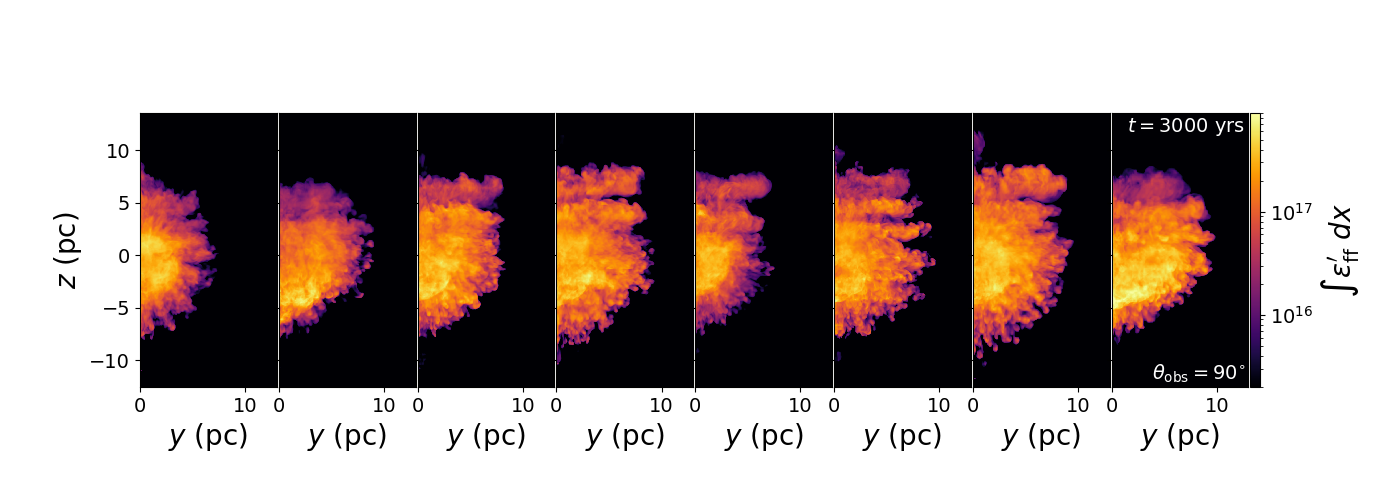}
    \includegraphics[width=1.0\linewidth,trim={1cm 2.6cm 0.5cm 2.8cm},clip]{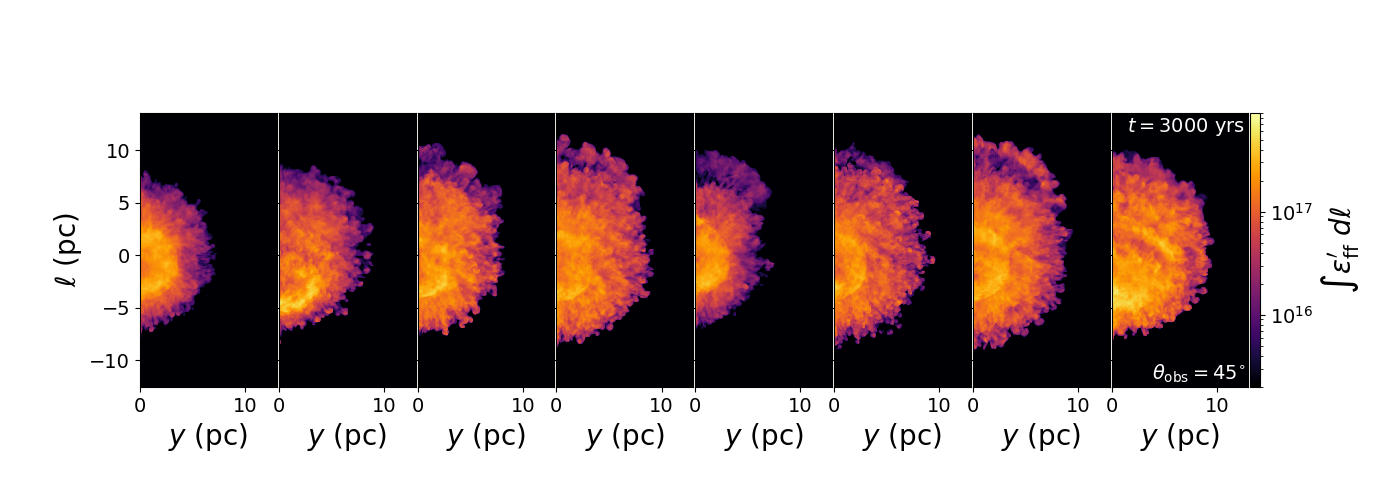}
    \includegraphics[width=1.0\linewidth,trim={1cm 2.8cm 0.5cm 4.55cm},clip]{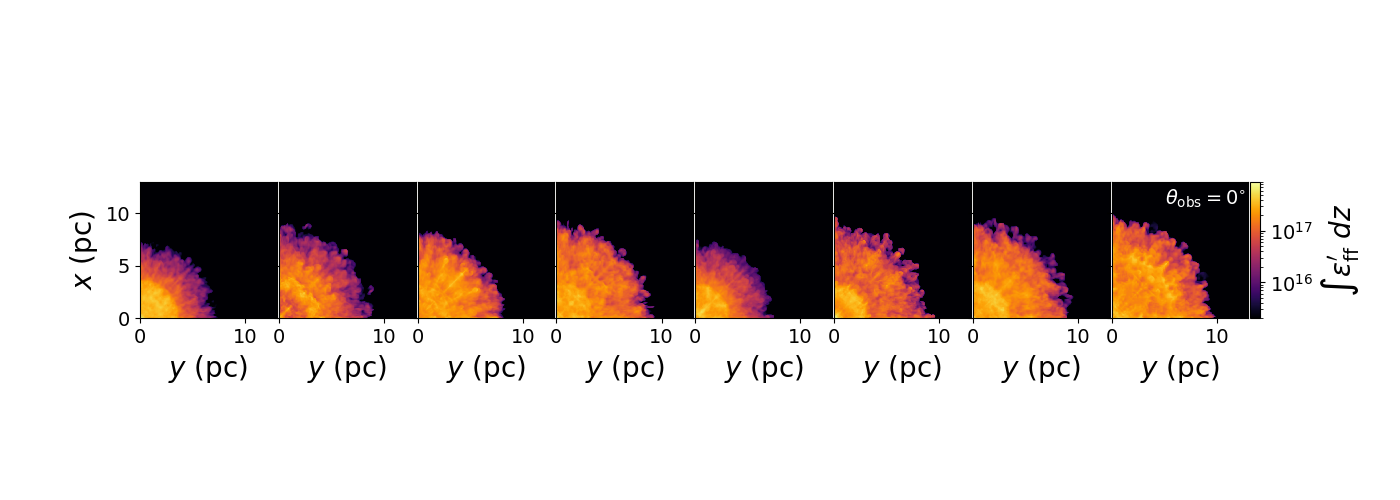}
    \caption{Ejecta mass fraction on the $x=0$ plane \textit{(top row)} and thermal X-ray emission proxy projected along the $x$-axis \textit{(second row)}, $x$-$z$-axis \textit{(third row)} direction, and $z$-axis \textit{(bottom row)}, all at $t=3000$ yrs.}
    \label{fig:chi_epsilon}
\end{figure*}

\begin{figure*}
\begin{center}
\begin{tabular}{ll}
  \includegraphics[height=0.6\textwidth,trim={2cm 0 0 0},clip]{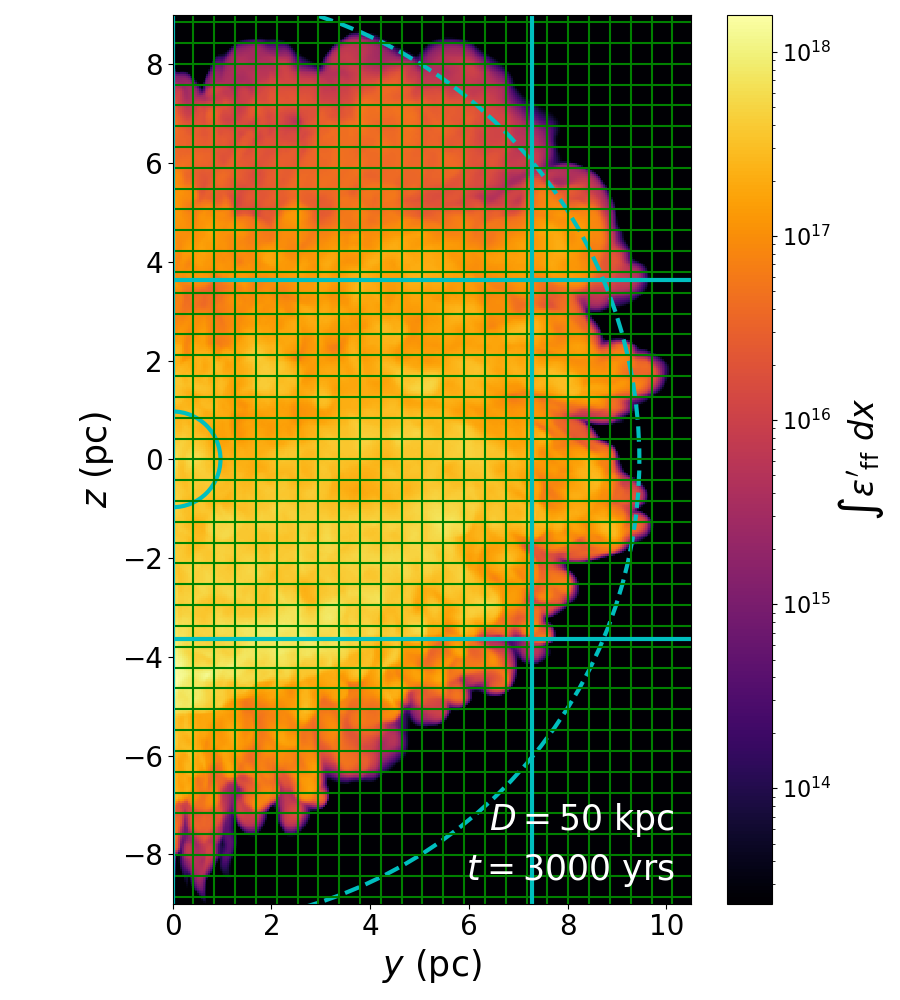} &
  \includegraphics[height=0.6\textwidth,trim={2cm 0 0cm 0},clip]{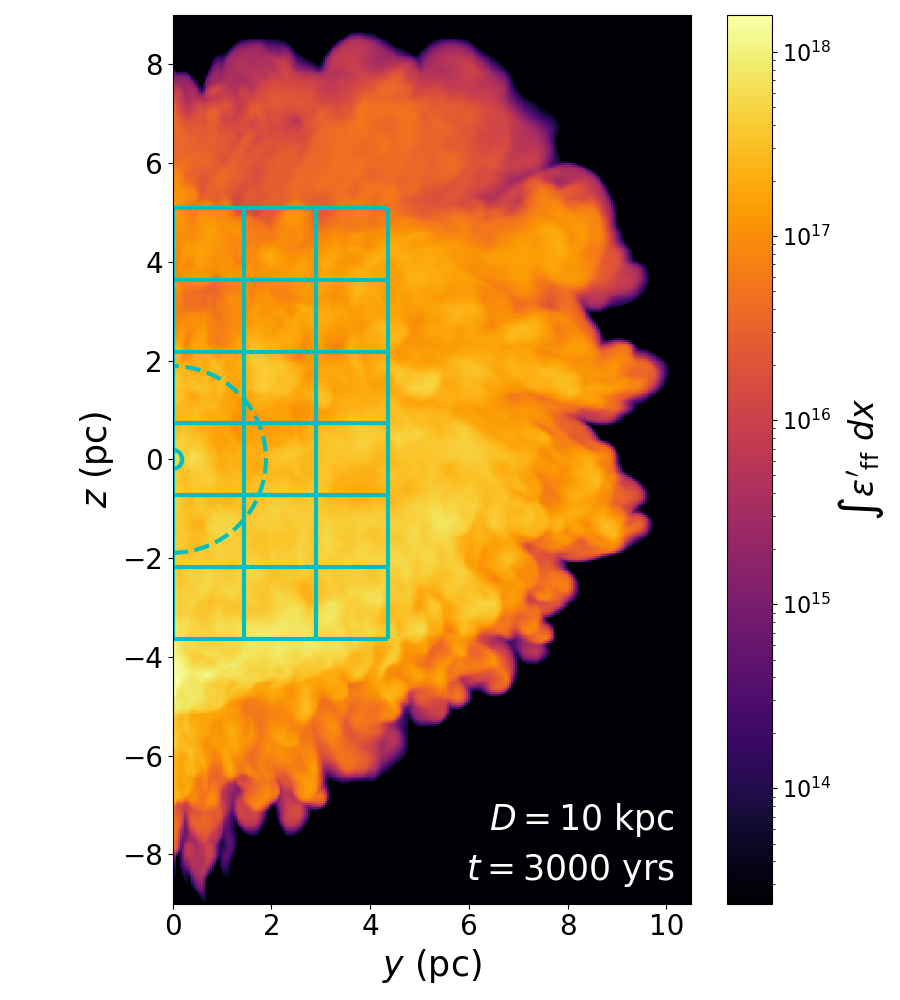}
\end{tabular}
\end{center}
\caption{Simulated observations of the $1.1+1$ $\msun$ direct detonation remnant at $t=3000$ yrs at a distance of 50 kpc \textit{(left)} and 10 kpc \textit{(right)}. The size of the XRISM pixels are shown as solid lines, with green lines corresponding to Xtend and thick blue lines to Resolve. In the right panel we do not show the Xtend pixels for readability. The solid and dashed circles show the FWHM and HPD of the Resolve PSF, which are similar to those of Xtend.}
\label{fig:xrism}
\end{figure*}

The free-free bremsstrahlung emission from shocked ejecta is given by
\be
\begin{aligned}
\varepsilon_{\rm ff} &= \frac{2^{5}\pi e^{6}}{3m_{e}c^{3}}\sqrt{\frac{2\pi}{3k_{B}m_{e}}}g_{\rm ff}(T_{e},\nu) T_{e}^{-1/2} \\ &\times \exp\left(-\frac{h\nu}{k_{B}T_{e}}\right)n_{e}\sum_{i}n_{i}Z_{i}^{2}~\frac{\rm ergs}{{\rm s}~{\rm cm}^{3}~{\rm Hz}} \label{eq:epsilon_ff}
\end{aligned}
\ee
\citep{2012A&ARv..20...49V}, for electron temperature $T_{e}$, Gaunt factor $g_{\rm ff}\approx 1$, frequency $\nu$, electron density $n_{e}$, and number density $n_{i}$ and atomic number $Z_{i}$ for each ion species. Determining $T_{e}$ would require an estimate of the ionization age of each gas parcel and the shock speed \citep[e.g.][]{2023ApJ...949...50R,2025ApJ...992...30I}, which our methods do not yield. \lp{Rather, we turn to observations of Ia SNRs, which yield an electron temperature of order 1 keV for young remnants \citep{2021PASJ...73..728Y}. This is insufficient to achieve full ionization, so we determine the electron density considering only electrons with a binding energy greater than 1 keV according to \citet{2008chcp.book.....L}.} We neglect the $\nu$ dependence and assume roughly isothermal electrons in the shocked gas, thereby focusing on the term in (\ref{eq:epsilon_ff}) which varies spatially:
\be
\varepsilon'_{\rm ff} = n_{e}\sum_{i}n_{i}Z_{i}^{2}.
\ee
As $n_{e}=\sum_{i}n_{i}N_{{\rm keV},i}$ and $n_{i}=X_{i}\rho/m_{i}$, where $m_{i}$ and $X_{i}$ are the mass and mass fraction of each ion species \lp{and $N_{{\rm keV},i}$ is the number of free electrons per ion at 1 keV,} this can be written as 
\be
\varepsilon'_{\rm ff} = \rho^{2}\left(\sum_{i}\frac{X_{i}N_{{\rm keV},i}}{m_{i}}\right)\left(\sum_{i}\frac{X_{i}Z_{i}^{2}}{m_{i}}\right).
\ee
Because the SNRs are optically thin, we choose as an emission measure the integral of this quantity along a line of sight. \lp{We have also computed $\varepsilon'_{\rm ff}$ under the assumption of full ionization, which overestimates the number of free electrons contributed by heavy elements (i.e.~iron). This did not produce qualitative differences in the results---iron dominates the emission in either case---so it appears that our results are not sensitive to our assumptions regarding ionization state of the plasma.}

As the data provided in \citet{boos_2024_10515767} gives the ejecta composition, we track the evolution of several elements (He, N, O, Si, S, and Fe) through the remnant phase using passive tracers to obtain the relevant $X_{i}$. The mass fraction of all ejecta vs ISM $X_{\rm ej}$ on the $x=0$ plane and the resulting emission measure integrated along the $x$-axis at $t=3000$ yrs are shown in \lp{the first and second rows of} Fig.~\ref{fig:chi_epsilon}. Many of the detonation models exhibit an ejecta distribution which tends to align with the symmetry axis. Triple detonations are the exception, in which the RS draws ISM into the remnant through the wake to reach the center of the remnant. The emission measure $\int\varepsilon'_{\rm ff}dx$ shows that all models display more asymmetry than the isolated WD remnant, exhibiting a flat top in the $+z$-direction. However, there are few distinguishable features among the models \lp{with the exception that the triple detonations are somewhat brighter in the southern hemisphere.} \lp{We also project along two additional viewing angles $\theta_{\rm obs}=45^{\circ}$ and $\theta_{\rm obs}=0^{\circ}$. Here it is no easier to distinguish between models, though the $z$-projections do highlight the smaller size of the double detonation remnants; they are also more centrally-concentrated, with much of the emission coming from within the bounce shock.} Thus, line emission may be necessary to differentiate between \lp{triple and quadruple detonations.}

In Fig.~\ref{fig:xrism}, we compare the capabilities of XRISM to one of our remnants ($1.1+1$ $\msun$, direct detonation) by overlaying the properties of both XRISM instruments on the emission measure. Here we show the size of the X-ray imager Xtend pixels as green lines, and of the spectrometer Resolve pixels as thick blue lines. The full width at half maximum (FWHM) of the point spread function (PSF) is shown as a solid circle, and the half-power diameter (HPD) of the PSF as a dashed circle. As the PSF is similar between the two instruments, we plot only the Resolve PSF properties. We consider a 3000-year-old remnant at a distance of 50 kpc (i.e.~LMC objects such as 0509-67.5 and 0519-69.0) as well as 10 kpc (i.e.~galactic SNRs such as G337.2-0.7 and G1.9+0.3). For readability, we do not show the Xtend pixels at 10 kpc. We see that Xtend is able to image the remnant at high spatial resolution at both distances. At a distance of 50 kpc, the remnant fills only a few Resolve pixels, whereas at 10 kpc the remnant exceeds the field of view of Resolve and would require multiple tilings to obtain spectra from all parts of the SNR.

\subsection{Elemental Composition} \label{sec:abundances}

\begin{figure*}[p!]
    \centering
    \includegraphics[width=0.81\linewidth,trim={0.9cm 0 0 0},clip]{figures/labels3.png}
    \includegraphics[width=1.0\linewidth,trim={1cm 2.9cm 0.5cm 2.8cm},clip]{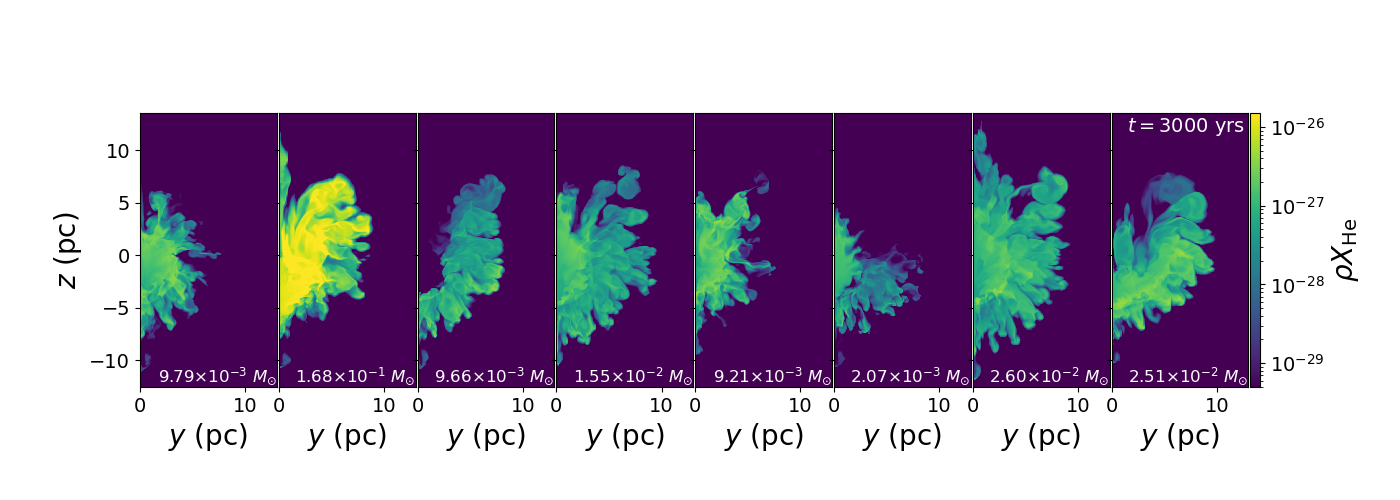}
    \includegraphics[width=1.0\linewidth,trim={1cm 2.9cm 0.5cm 2.8cm},clip]{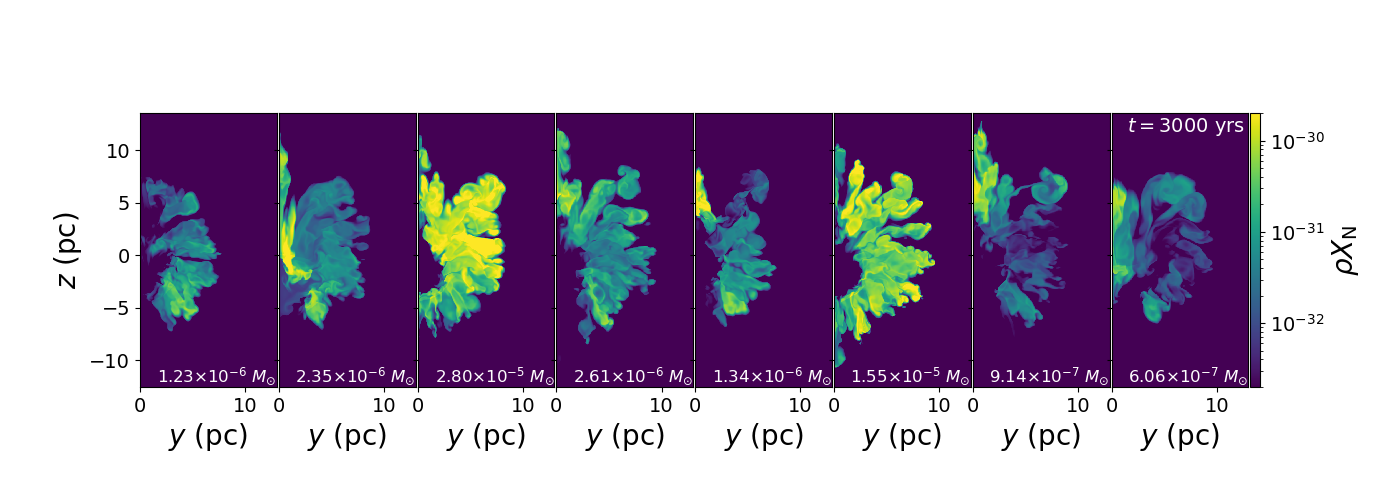}
    \includegraphics[width=1.0\linewidth,trim={1cm 2.9cm 0.5cm 2.8cm},clip]{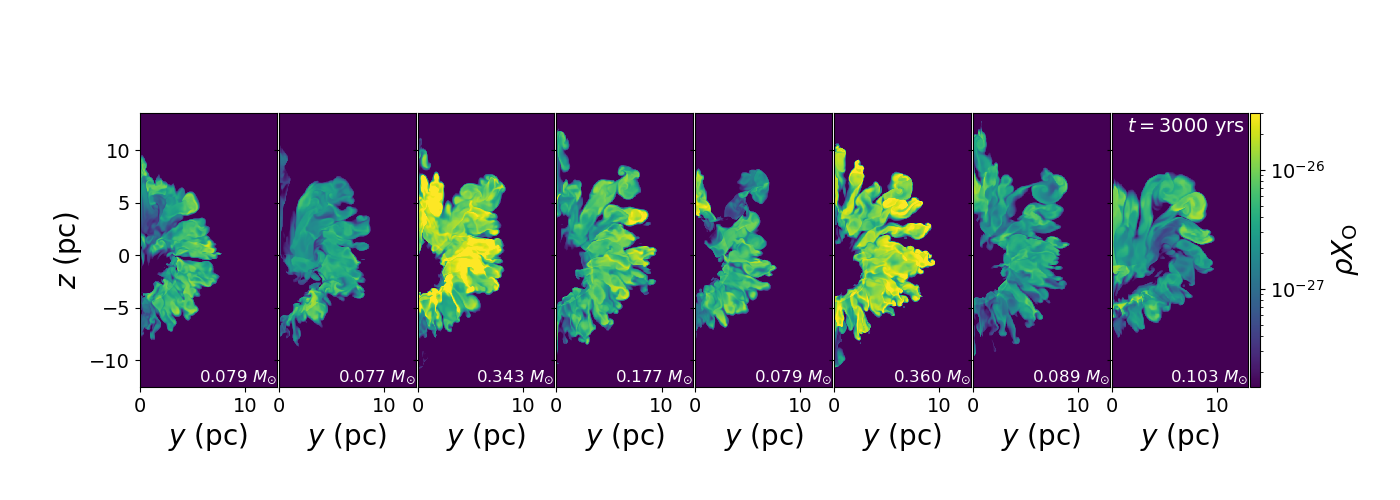}
    \includegraphics[width=1.0\linewidth,trim={1cm 2.9cm 0.5cm 2.8cm},clip]{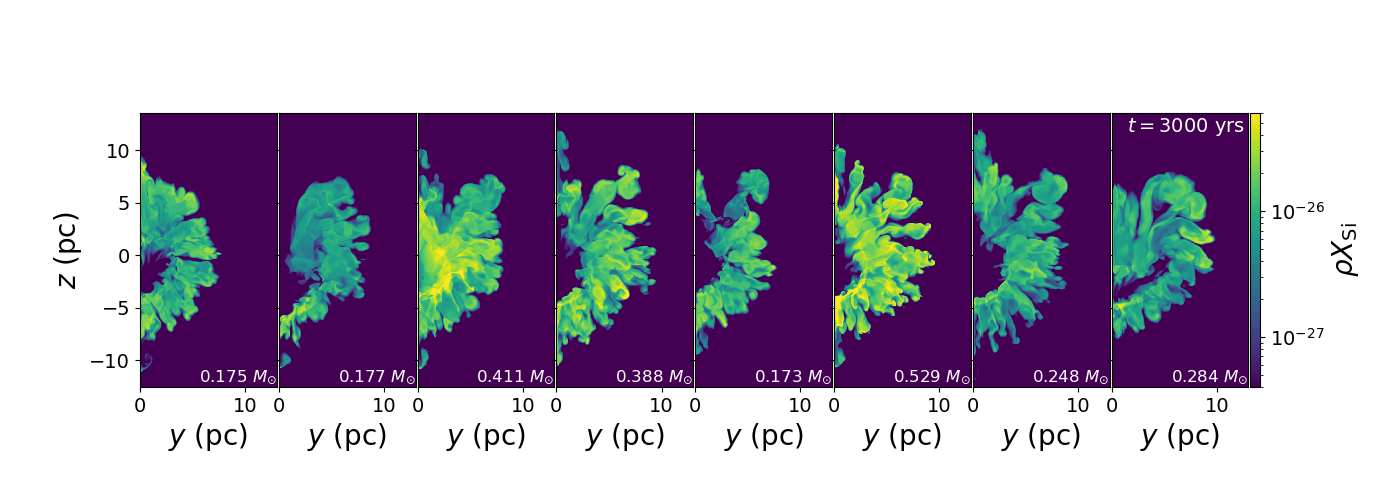}
    \includegraphics[width=1.0\linewidth,trim={1cm 1cm 0.5cm 2.8cm},clip]{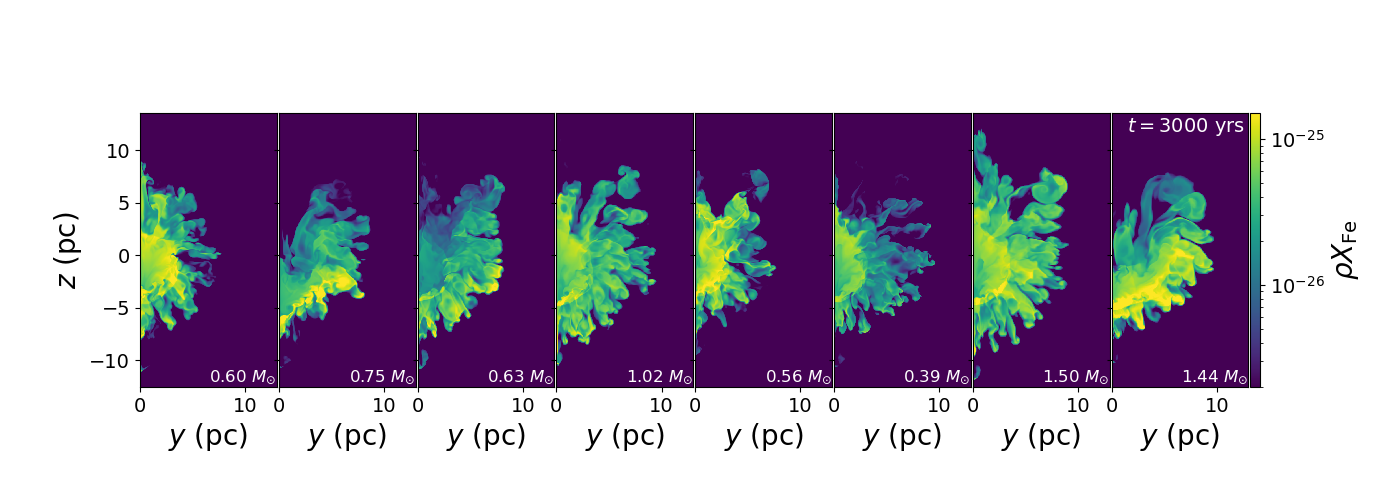}
    \caption{Elemental mass density slices on the $x=0$ plane at $t=3000$ yrs. \lp{Each panel is labeled with the total mass of that element present in the remnant.}}
    \label{fig:tracergrid}
\end{figure*}

\begin{figure*}[p!]
    \centering
    \includegraphics[width=0.81\linewidth,trim={0.9cm 0 0 0},clip]{figures/labels3.png}
    \includegraphics[width=1.0\linewidth,trim={1cm 2.9cm 0.5cm 2.8cm},clip]{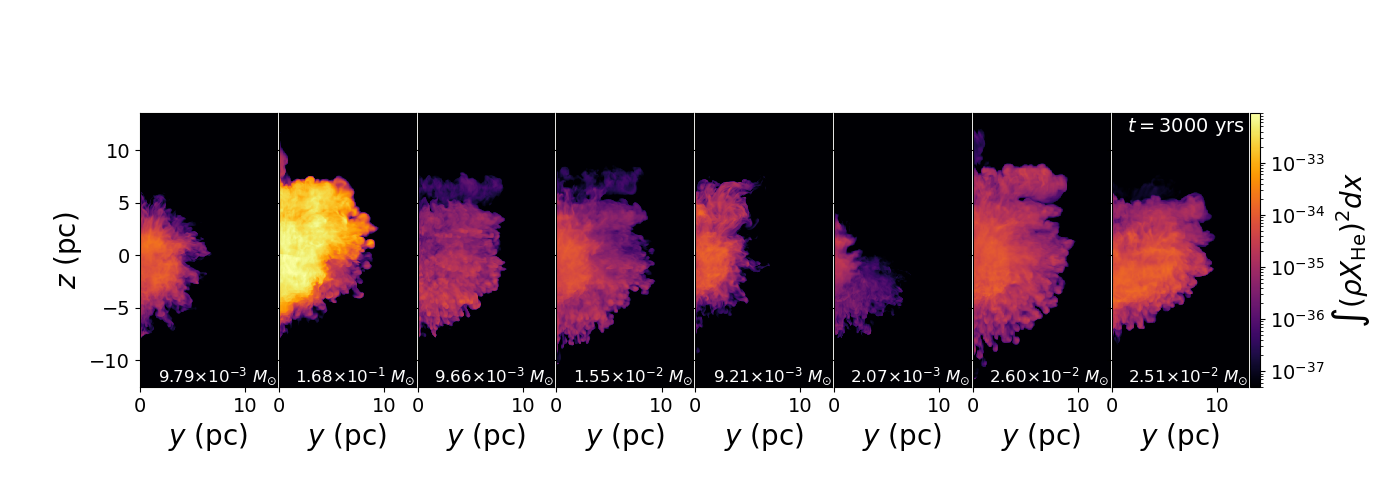}
    \includegraphics[width=1.0\linewidth,trim={1cm 2.9cm 0.5cm 2.8cm},clip]{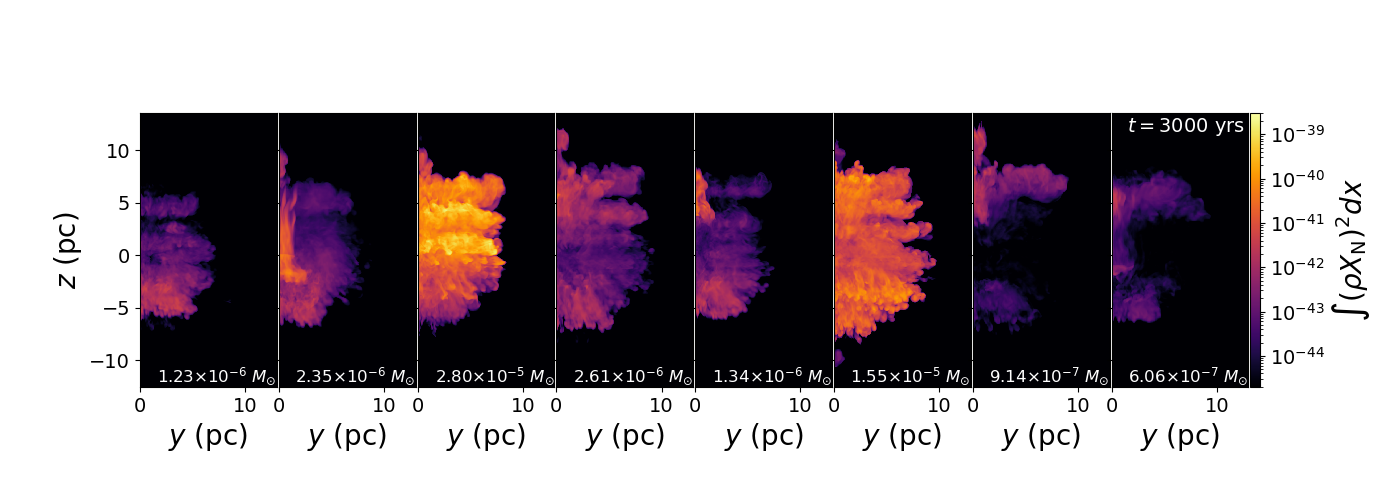}
    \includegraphics[width=1.0\linewidth,trim={1cm 2.9cm 0.5cm 2.8cm},clip]{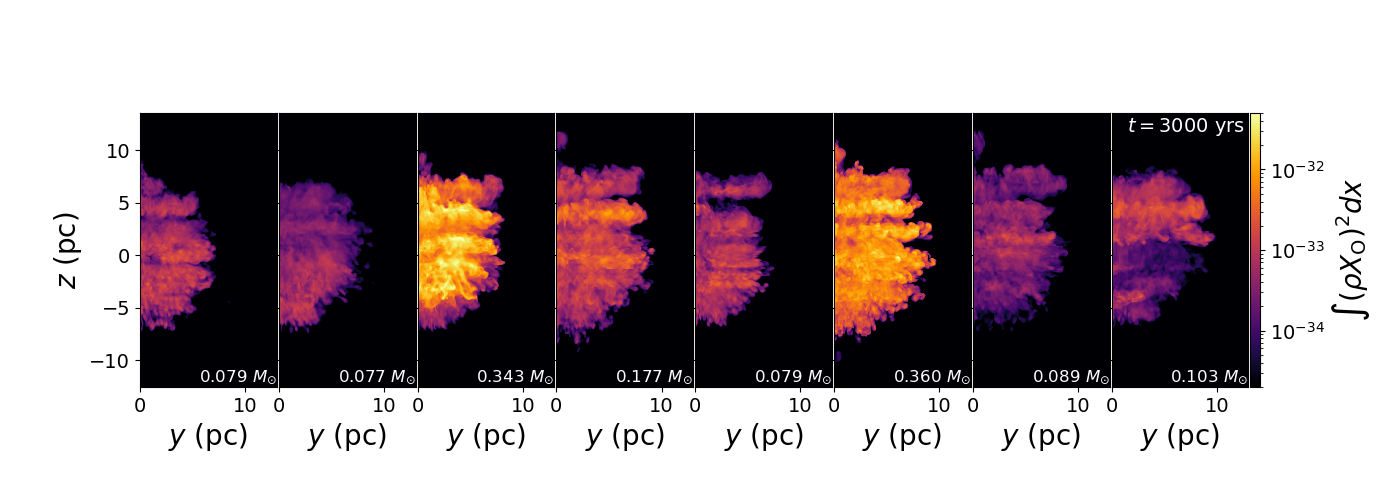}
    \includegraphics[width=1.0\linewidth,trim={1cm 2.9cm 0.5cm 2.8cm},clip]{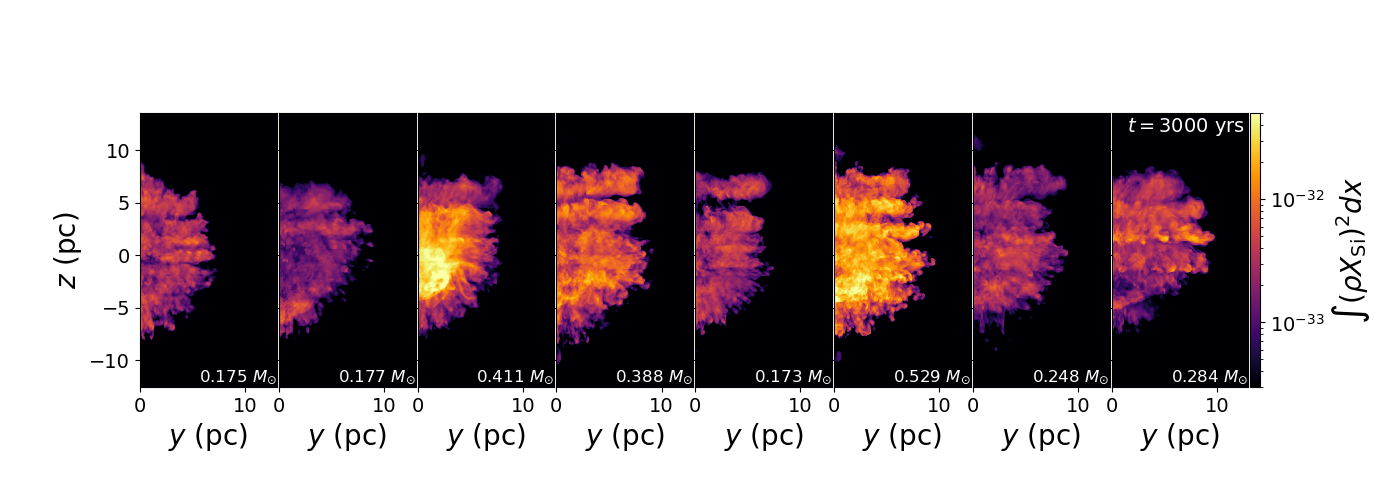}
    \includegraphics[width=1.0\linewidth,trim={1cm 1cm 0.5cm 2.8cm},clip]{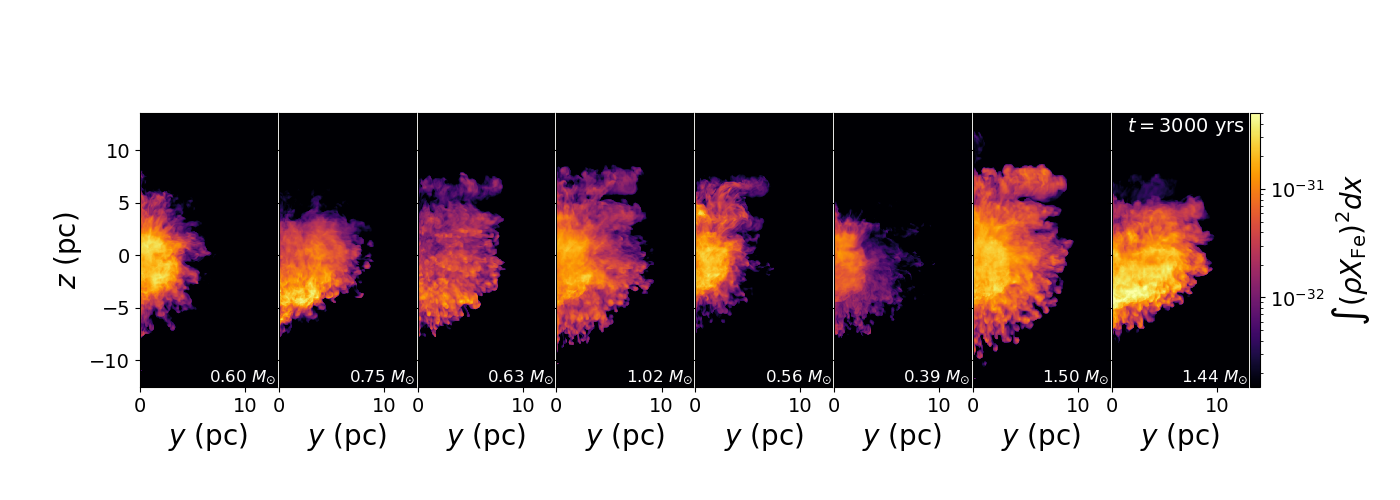}
    \caption{Squared elemental mass density projections along the $x$-axis at $t=3000$ yrs. \lp{Each panel is labeled with the total mass of that element present in the remnant.}}
    \label{fig:rhotracergrid}
\end{figure*}

As X-ray spectrometers are adept at discerning alpha-process elements \citep{2012A&ARv..20...49V}, we investigate here whether the line emission from individual elements may provide a better probe of detonation physics than $\varepsilon'_{\rm ff}$. The nucleosynthetic yields of each explosion---listed in BTS24---vary with the progenitor masses, as the yields are sensitive to the densities encountered by the detonation wave. However, their distributions are greatly influenced by the dynamics of the SNR presented in section \ref{sec:dynamics}.

We first consider planar slices of the \lp{densities} of several elements (He, N, O, Si, and Fe), shown in Fig.~\ref{fig:tracergrid}, at $t=3000$ yrs. By this point, all $^{56}$Ni and $^{56}$Co has decayed to Fe. We also traced sulfur, but omit it in the plots for space and because it is generally similar to silicon (but roughly half as abundant). Here there are clear differences between the elements across the suite of models, particularly their abundance in the core of the SNR and in the wake. For example, most remnants do not contain significant central concentrations of silicon, with the exceptions of the $1+0.7$ $\msun$ and $1.1+1$ $\msun$ (direct detonation) models. Both the 0.7 and 0.8 $\msun$ donor detonations produce Si to low velocities, so this discrepancy at late times can be attributed to differing reverse shock dynamics. On the other hand, $1+0.7$ $\msun$ is the only model which does not contain significant helium in its core or wake. Modest changes in WD masses can have a large impact, with the $1+0.7$ and $1+0.9$ $\msun$ models visibly differing across all elements traced. 
%A large fraction of nitrogen is found in the wake in all triple detonations as well as the surviving donor case. 
%Iron plays an important role in SNR tomography, as it can be used to identify the location of the reverse shock \citep[e.g.][]{2021ApJ...923..141L}.

\lp{We take the quantity $(\rho X_{i})^{2}$ projected along a line of sight as a proxy for X-ray line emission, shown in Fig.~\ref{fig:rhotracergrid}. Here many of the discrepancies present in the slice plots are eliminated, though several remain.} The mass of the primary appears to play a role in the morphology: the $0.85+0.8$ $\msun$ SNR displays a unique helium distribution, whereas the $1.1$ $\msun$ primaries have a characteristic nitrogen distribution regardless of the ignition point. \lp{That said, the total mass of nitrogen in all eight remnants is multiple orders of magnitude smaller than that of other species, as shown by the labels on each panel in Figs.~\ref{fig:tracergrid} and \ref{fig:rhotracergrid}.} \lp{Though iron exhibits fewer discrepancies between models than light elements such as He and N, its high mass fraction and atomic number make observations of such discrepancies far easier. The triple detonations show exhibit more iron line emission in the southern hemisphere than double or quadruple detonations---similar to the thermal emission---though the 0.85+0.8 $\msun$ model is an exception to this.} Comparing these distributions to the XRISM coverage shown in Fig.~\ref{fig:xrism}, we see that XRISM-Resolve has sufficient spatial resolution to probe the line emission in galactic remnants. LMC remnants subtend only a few Resolve pixels, though this may be sufficient to obtain meaningful spectra of remnants of this age. While nitrogen lines can be identified in X-ray spectroscopy of SNRs \citep[e.g.][]{2021ApJ...916...41S}, helium features are difficult to discern and may be restricted to other bands.

The total masses of elements contained in the SNRs (listed in Table \ref{tab:detmodels}) vary substantially according to the nature of the donor. For example, a portion of the helium in the He WD donor escapes detonation, leaving its remnant with the largest helium mass by an order of magnitude. In addition, donor mass tends to scale with iron mass and inversely to oxygen mass. Though a fraction of the nitrogen originates from the pre-detonation shells, it can be produced in significant quantities in the detonation of intermediate-mass donors. This can be seen in the $1+0.7$ $\msun$ and $0.85+0.8$ $\msun$ models in the second row of Fig.~\ref{fig:rhotracergrid}. This is also seen, to a lesser degree, in silicon. In general, \lp{while He and N exhibit the greatest variation between models, thermal and iron-line emission provide the largest observable discrepancies between detonations.}

\section{Discussion and Conclusions} \label{sec:discussion}

We have simulated the evolution of the post-detonation ejecta models presented in BTS24 through the remnant phase, yielding their physical structure as well as estimates of their X-ray emission. These models span a parameter space of binary masses and donor ignition points (or lack thereof), facilitating direct comparisons. Prior to the remnant phase, the ejecta contains two main features which will impact its subsequent evolution: 
\begin{enumerate}
    \item Hydrodynamical interactions between the ejecta and donor carve out a low-density wake bounded by an overdensity at the bow shock. Even if the donor detonates, these features form prior to that detonation.
    \item A donor detonation injects a shell of dense ejecta at low velocity with different nucleosynthetic yields than that of the primary. This shell is larger in extent in the case of triple detonations due to the shorter time delay between detonations.
\end{enumerate}
%Additionally, the thermodynamics of the wake differ markedly from those of the unshocked ejecta. This has little effect on the remnant phase---all ejecta eventually reaches negligible temperature due to adiabatic expansion prior to crossing the reverse shock---but may affect the early light curve (Kumar et al., submitted to ApJ). The unique composition of the wake may also impact emission line profiles during the nebular phase \citep{2025arXiv250706412S}.

The features listed above have direct consequences for the shock surfaces. The forward shock initially maintains a shape similar to that of the post-detonation ejecta. Aside from complications within the wake, the overall size of the FS scales with the mass of the primary, as all of the high-velocity ejecta originated from the primary. All models less the isolated WD exhibit a protrusion in the FS at the location of the shock cone, \lp{though cosmic rays have been proposed as a mechanism to create similar ``ear''-like structures \citep[e.g.][]{2005ApJ...634..376W,2025MNRAS.543.2791Y}.} After several centuries, the forward shock slows to a velocity comparable to that of the shell of donor ejecta. This bolsters the FS, and at late times remnants of triple and quadruple detonations are larger than those of double detonations in all cases. This can be quantified by fitting each remnant to a sphere, demonstrating that at $t=3000$ yrs the $1+0.7$ $\msun$ quadruple-detonation SNR is larger in diameter by 7\% than its double-detonation counterpart. Additionally, these fits reveal that the geometric centers of all 3000-year-old remnants are offset from the location of the supernova by 0.5--1 pc due to the fact that the primary is detonated off-center. Observationally, the forward shock may be traced using H$\alpha$ emission, and the shock speed by the width of H$\alpha$ line \citep{1982ApJ...261..473T, 2008ApJ...689.1089V, 2013A&A...558A..25M, 2018ApJ...862..148H}.

The reverse shock behaves similarly to the FS at early times, but is uniquely affected by the wake \citep{2025ApJ...982...60P}. In quadruple detonations, the roughly-spherical shell of donor ejecta tends to slow and sphericize the RS. However, in triple detonations the asymmetrical shell provides a highway of low-density ejecta which allows the RS to quickly propagate into the SNR (third row of Fig.~\ref{fig:rhogrid}). This leads to a large deformation of the RS shape prior to convergence (fourth row of Fig.~\ref{fig:rhogrid}). A similar phenomenon has been documented in the case of a surviving donor in a variety of studies \citep{2012ApJ...745...75G, 2016ApJ...833...62G, 2022ApJ...930...92F, 2025ApJ...982...60P}. We find that the effect is in fact more pronounced in triple detonations. Additionally, ISM material may be drawn into the wake to an extent such that it reaches the center of the SNR (Fig.~\ref{fig:chi_epsilon}). The resulting dearth of heavy elements within the wake may be responsible for the region of SNR 0519-69.0 which emits in H$\alpha$ but not in X-rays \citep{2021ApJ...923..141L,2022ApJ...935...78W}, though we have yet to conduct an in-depth investigation of any particular SNR.

Though the forward shocks eventually become spherical, the composition of the SNRs lead to asymmetries in both the thermal X-rays and in the line emission. Estimates of the continuum thermal emission show that the presence of the wake plays a large role, in agreement with previous studies \citep{2022ApJ...930...92F, 2025ApJ...982...60P}. We find that the thermal emission is fairly similar across all detonation mechanisms (Fig.~\ref{fig:chi_epsilon}) \lp{but is skewed toward the southern hemisphere for triple detonations. It also serves to reveal the relatively smaller size of double detonation remnants.} However, the distributions of individual species can differ substantially even between remnants with similar progenitor binaries. For example, the $1+0.7$ $\msun$ and $1+0.9$ $\msun$ models differ qualitatively for all elements traced (Fig.~\ref{fig:tracergrid}) when considering slices of their distributions at the $x=0$ plane. Many of these distinguishing features are eliminated when we take a projection of each \lp{$(\rho X_{i})^{2}$} along a line of sight, which provides a better proxy for line emission. \lp{However, the iron line emission displays a brighter southern hemisphere for triple detonations in a similar manner to the continuum thermal emission, as iron is the primary contributor to thermal emission.} Discrepancies also persist for the lighter elements He and N, which are not produced in great quantities during the supernova but trace the disparate remnant dynamics discussed above. Nitrogen lines \lp{are rarely} identified in supernova remnants \citep[\lp{with notable exceptions such as} SN1987A with XMM-Newton,][]{2021ApJ...916...41S}, \lp{limiting their use as a probe of explosion properties}. In the single-degenerate case, the circumstellar medium may contain significant nitrogen \citep{2021ApJ...915...42K}---potentially masking nitrogen originating from the WD---but this is not expected in double-degenerate binaries.

These findings are promising considering the recent use of XRISM to probe the structures of core-collapse SNRs \citep{2025arXiv250507479X, 2025arXiv250504691V, 2025arXiv250403268B, 2025arXiv250323640S, 2025arXiv250403223G}. As shown in Fig.~\ref{fig:xrism}, for galactic remnants the XRISM-Resolve spectrometer is able to obtain spectra with a spatial resolution of $<2$ pc. Comparison with Fig.~\ref{fig:rhotracergrid} demonstrates that this resolution is sufficient to discriminate between the detonation models we consider here. For LMC remnants, each Resolve pixel covers a significant fraction of the SNR, but may still be able to obtain meaningful spectra. Additionally, the total masses of these species (Table \ref{tab:detmodels}) are functions of the progenitor binary giving an additional constraint on the physics of the explosion. The helium distribution may be more challenging to observe, as helium lines are not identified in the X-ray band in SNRs. However, it has been observed in the optical band in remnants such as the Crab nebula \citep[e.g.][]{1982ApJ...258...11H} and Cassiopeia A \citep{2008Sci...320.1195K}.

\lp{\citet{2025arXiv251018800F} recently carried out two simulations of the remnants of a WD binary following both a double and a quadruple detonation, based on the results of \citet{2022MNRAS.517.5260P}. The ISM density we have assumed here is $\approx$4 times higher than that of \citet{2025arXiv251018800F}, meaning that their integration time of 1000 yrs corresponds to an age of $1000\times 4^{1/3}\approx 625$ yrs when comparing to the remnants presented here. The results of our work generally agree with theirs up to this epoch, though they also found (among other things) that an excess of unburnt carbon in the wake could be a feature unique to double detonations. This may also be present in the remnants of the BTS24 detonations---we did not trace carbon due to its low mass fraction---though the composition of the progenitor WDs of \citet{2022MNRAS.517.5260P} differ from those of BTS24.}

\lp{Here we have assumed a uniform ISM with a density based on that measured for one particular SNR. As such, the epochs we report here are dependent on this density, with quantities such as the time needed for the FS to reach spherical symmetry, the size of the remnant relative to the XRISM FOV, etc.~are expected to scale as $\rho_{\rm ISM}^{-1/3}$. Additionally, phenomena such as an ISM density gradient or a collision with a molecular cloud create a dipole effect similar to that of the off-center detonation of the primary. A bulk velocity of the SNR relative to the ISM has been found to create a similar result \citep{2022ApJ...930...92F}, so these effects may be difficult to differentiate from one another. Supernovae within planetary nebulae have received relatively little study but are also expected to be bipolar in shape due to equatorial outflows during a common envelope event \citep[e.g.][]{2018ApJ...860...19G}.}

\citet{2011ApJ...732..114L} and \citet{2025arXiv250513671L} have found that SNIa remnants possess more mirror symmetry than core-collapse SNRs. Though the remnants we have presented here are consistent with these findings, our simulations are based on 2D detonation calculations with inherent azimuthal symmetry. In particular, the ignition point of the primary was placed on the axis of symmetry, which may not be the case in reality. For off-axis ignition, the asymmetry created by the off-center detonation of the donor would not be coaxial with that created by the donor interaction and detonation. In future work, it will be necessary to model the detonations in 3D to facilitate off-axis ignition. The effects that may arise due to the helium shell ejecta in the remnant phase, which were not considered in this work, should also be examined. These effects are expected to be small due to the low mass of the shells, though \citet{2022ApJ...930...92F} showed that at $t\sim 100$ yrs the primary shell detonation produces a protrusion opposite to its ignition point. \lp{Ignition of the helium shell is assumed to occur at Roche lobe overflow in BTS24, which sets the orbital separation. In reality, the dynamics of the ignition are complex, so the actual separation at ignition may differ from this value. 
For quadruple detonations, the time needed for the blast wave to traverse the vacuum between the white dwarfs comprises only a fraction of the delay time between the detonations, so we do not expect this to be a large source of error; for triple detonations, it may be more significant.} Because the time delay between the detonations appears to have a large impact on remnant morphology, it is also important to determine in which cases one shell detonation is sufficient to ignite the other shell \citep[e.g.][]{2021MNRAS.503.4734P}. 

\lp{A related consideration is that the donor's velocity ($\sim$2,000 km/s) is expected to affect the structure of the primary ejecta at comparable velocities. This is also a significant fraction of the velocity of the donor's own ejecta ($\approx$10,000 km/s in homology), so the motion of the donor during the time delay between detonations may result in asymmetry in the shell of donor ejecta.}

%Additional work is also needed to explore the mapping between progenitor binaries and detonation mechanisms, as the regions of parameter space which correspond to double, triple, and quadruple detonations are not well-defined.

\begin{acknowledgments}
We thank Ken Shen for useful discussions regarding the detonation models, Soham Mandal for helpful pointers on the use of Sprout, and Joseph Farah for discussions on SNR composition. We also acknowledge a thoughtful report from the anonymous referee. This research benefited from interactions with a variety of researchers that were funded by the Gordon and Betty Moore Foundation through Grant GBMF5076. Computational resources for this work were provided by the Expanse supercomputer at the San Diego Supercomputer Center through allocation PHY250020 from the Advanced Cyberinfrastructure Coordination Ecosystem: Services \& Support (ACCESS) program \citep{access}. ACCESS is supported by National Science Foundation grants \#2138259, \#2138286, \#2138307, \#2137603, and \#2138296. This research was supported in part by grant NSF PHY-2309135 to the Kavli Institute for Theoretical Physics (KITP). LJP is supported by a grant from the Simons Foundation (216179, LB) as well as a grant from the NASA Astrophysics Theory Program (ATP-80NSSC22K0725). The Flatiron Institute is supported by the Simons Foundation. We use the Matplotlib \citep{Hunter:2007} and SciPy \citep{2020SciPy-NMeth} software packages for the generation of plots in this paper.
%Use was made of computational facilities purchased with funds from the National Science Foundation (CNS-1725797) and administered by the Center for Scientific Computing (CSC). The CSC is supported by the California NanoSystems Institute and the Materials Research Science and Engineering Center (MRSEC; NSF DMR 2308708) at UC Santa Barbara.
\end{acknowledgments}

\software{Matplotlib \citep{Hunter:2007},
          NumPy \citep{harris2020arrayNUMPY},
          SciPy \citep{2020SciPy-NMeth},
          Sprout \citep{sprout}
          }

%\newpage

%\appendix


\bibliography{references}{}

@ARTICLE{2025arXiv250323640S,
       author = {{Suzuki}, Shunsuke and {Sonoda}, Haruto and {Sakai}, Yusuke and {Ohshiro}, Yuken and {Yamada}, Shinya and {Agarwal}, Manan and {Katsuda}, Satoru and {Yamaguchi}, Hiroya},
        title = "{Dynamics of the intermediate-mass-element ejecta in the Supernova Remnant Cassiopeia A studied with XRISM}",
      journal = {arXiv e-prints},
         year = 2025,
        month = mar,
          eid = {arXiv:2503.23640},
        pages = {arXiv:2503.23640},
          doi = {10.48550/arXiv.2503.23640},
archivePrefix = {arXiv},
       eprint = {2503.23640},
 primaryClass = {astro-ph.HE},
       adsurl = {https://ui.adsabs.harvard.edu/abs/2025arXiv250323640S}
}

@ARTICLE{2025arXiv250403268B,
       author = {{Bamba}, Aya and {Agarwal}, Manan and {Vink}, Jacco and {Plucinsky}, Paul and {Terada}, Yukikatsu and {Behar}, Ehud and {Katsuda}, Satoru and {Mori}, Koji and {Sawada}, Makoto and {Matsumoto}, Hironori and {Corrales}, Lia and {Foster}, Adam and {Fujimoto}, Shin-ichiro and {Gu}, Liyi and {Ichikawa}, Kazuhiro and {Matsunaga}, Kai and {Mizuno}, Tsunefumi and {Murakami}, Hiroshi and {Nakajima}, Hiroshi and {Sato}, Toshiki and {Sonoda}, Haruto and {Suzuki}, Shunsuke and {Tateishi}, Dai and {Uchida}, Hiroyuki and {Ichihashi}, Masahiro and {Nobukawa}, Kumiko and {Orlando}, Salvatore},
        title = "{Measuring the asymmetric expansion of the Fe ejecta of Cassiopeia A with XRISM/Resolve}",
      journal = {arXiv e-prints},
         year = 2025,
        month = apr,
          eid = {arXiv:2504.03268},
        pages = {arXiv:2504.03268},
          doi = {10.48550/arXiv.2504.03268},
archivePrefix = {arXiv},
       eprint = {2504.03268},
 primaryClass = {astro-ph.HE},
       adsurl = {https://ui.adsabs.harvard.edu/abs/2025arXiv250403268B}
}

@ARTICLE{2025arXiv250504691V,
       author = {{Vink}, Jacco and {Agarwal}, Manan and {Bamba}, Aya and {Gu}, Liyi and {Plucinsky}, Paul and {Behar}, Ehud and {Corrales}, Lia and {Foster}, Adam and {Fujimoto}, Shin-ichiro and {Ichihashi}, Masahiro and {Ichikawa}, Kazuhiro and {Katsuda}, Satoru and {Matsumoto}, Hironori and {Matsunaga}, Kai and {Mizuno}, Tsunefumi and {Mori}, Koji and {Murakami}, Hiroshi and {Nakajima}, Hiroshi and {Sato}, Toshiki and {Sawada}, Makoto and {Sonoda}, Haruto and {Suzuki}, Shunsuke and {Tateishi}, Dai and {Terada}, Yukikatsu and {Uchida}, Hiroyuki},
        title = "{Mapping Cassiopeia A's silicon/sulfur Doppler velocities with XRISM-Resolve}",
      journal = {arXiv e-prints},
         year = 2025,
        month = may,
          eid = {arXiv:2505.04691},
        pages = {arXiv:2505.04691},
          doi = {10.48550/arXiv.2505.04691},
archivePrefix = {arXiv},
       eprint = {2505.04691},
 primaryClass = {astro-ph.HE},
       adsurl = {https://ui.adsabs.harvard.edu/abs/2025arXiv250504691V}
}

@ARTICLE{2025arXiv250403223G,
       author = {{Gu}, Liyi and {Yamaguchi}, Hiroya and {Foster}, Adam and {Katsuda}, Satoru and {Uchida}, Hiroyuki and {Sawada}, Makoto and {Porter}, Frederick Scott and {Williams}, Brian J. and {Petre}, Robert and {Bamba}, Aya and {Terada}, Yukikatsu and {Agarwal}, Manan and {Decourchelle}, Anne and {Guainazzi}, Matteo and {Kelley}, Richard and {Kilbourne}, Caroline and {Loewenstein}, Michael and {Matsumoto}, Hironori and {Miller}, Eric D. and {Ohshiro}, Yuken and {Plucinsky}, Paul and {Suzuki}, Hiromasa and {Tashiro}, Makoto and {Vink}, Jacco and {Ezoe}, Yuichiro and {Behar}, Ehud and {Smith}, Randall},
        title = "{Evidence for Charge Exchange Emission in Supernova Remnant N132D from XRISM/Resolve Observations}",
      journal = {arXiv e-prints},
         year = 2025,
        month = apr,
          eid = {arXiv:2504.03223},
        pages = {arXiv:2504.03223},
          doi = {10.48550/arXiv.2504.03223},
archivePrefix = {arXiv},
       eprint = {2504.03223},
 primaryClass = {astro-ph.HE},
       adsurl = {https://ui.adsabs.harvard.edu/abs/2025arXiv250403223G}
}

@ARTICLE{2025arXiv250507479X,
       author = {{XRISM Collaboration}},
        title = "{Thermal and Kinematic Properties of Ejecta in SN1987A revealed by XRISM}",
      journal = {arXiv e-prints},
         year = 2025,
        month = may,
          eid = {arXiv:2505.07479},
        pages = {arXiv:2505.07479},
archivePrefix = {arXiv},
       eprint = {2505.07479},
 primaryClass = {astro-ph.HE},
       adsurl = {https://ui.adsabs.harvard.edu/abs/2025arXiv250507479X}
}

@ARTICLE{2010ApJ...709L..64G,
       author = {{Guillochon}, James and {Dan}, Marius and {Ramirez-Ruiz}, Enrico and {Rosswog}, Stephan},
        title = "{Surface Detonations in Double Degenerate Binary Systems Triggered by Accretion Stream Instabilities}",
      journal = {\apjl},
         year = 2010,
        month = jan,
       volume = {709},
       number = {1},
        pages = {L64-L69},
          doi = {10.1088/2041-8205/709/1/L64},
archivePrefix = {arXiv},
       eprint = {0911.0416},
 primaryClass = {astro-ph.HE},
       adsurl = {https://ui.adsabs.harvard.edu/abs/2010ApJ...709L..64G}
}

@ARTICLE{2021MNRAS.503.4734P,
       author = {{Pakmor}, R. and {Zenati}, Y. and {Perets}, H.~B. and {Toonen}, S.},
        title = "{Thermonuclear explosion of a massive hybrid HeCO white dwarf triggered by a He detonation on a companion}",
      journal = {\mnras},
         year = 2021,
        month = jun,
       volume = {503},
       number = {4},
        pages = {4734-4747},
          doi = {10.1093/mnras/stab686},
archivePrefix = {arXiv},
       eprint = {2103.06277},
 primaryClass = {astro-ph.SR},
       adsurl = {https://ui.adsabs.harvard.edu/abs/2021MNRAS.503.4734P}
}

@ARTICLE{2005ApJ...634..376W,
       author = {{Warren}, Jessica S. and {Hughes}, John P. and {Badenes}, Carles and {Ghavamian}, Parviz and {McKee}, Christopher F. and {Moffett}, David and {Plucinsky}, Paul P. and {Rakowski}, Cara and {Reynoso}, Estela and {Slane}, Patrick},
        title = "{Cosmic-Ray Acceleration at the Forward Shock in Tycho's Supernova Remnant: Evidence from Chandra X-Ray Observations}",
      journal = {\apj},
         year = 2005,
        month = nov,
       volume = {634},
       number = {1},
        pages = {376-389},
          doi = {10.1086/496941},
archivePrefix = {arXiv},
       eprint = {astro-ph/0507478},
 primaryClass = {astro-ph},
       adsurl = {https://ui.adsabs.harvard.edu/abs/2005ApJ...634..376W}
}

@ARTICLE{2021PASJ...73..728Y,
       author = {{Yamauchi}, Shigeo and {Nobukawa}, Masayoshi and {Koyama}, Katsuji},
        title = "{A systematic comparison of ionization temperatures between ionizing and recombining plasmas in supernova remnants}",
      journal = {\pasj},
         year = 2021,
        month = jun,
       volume = {73},
       number = {3},
        pages = {728-734},
          doi = {10.1093/pasj/psab033},
archivePrefix = {arXiv},
       eprint = {2104.02375},
 primaryClass = {astro-ph.HE},
       adsurl = {https://ui.adsabs.harvard.edu/abs/2021PASJ...73..728Y}
}

@ARTICLE{2025ApJ...992...30I,
       author = {{Ichihashi}, Masahiro and {Bamba}, Aya and {Tateishi}, Dai and {Hagino}, Kouichi and {Katsuda}, Satoru and {Uchida}, Hiroyuki and {Suzuki}, Hiromasa and {Yamazaki}, Ryo and {Ohira}, Yutaka},
        title = "{The Electron Temperature Distribution and the High Ionization Just behind the Shock in the Cygnus Loop}",
      journal = {\apj},
         year = 2025,
        month = oct,
       volume = {992},
       number = {1},
          eid = {30},
        pages = {30},
          doi = {10.3847/1538-4357/ae058c},
archivePrefix = {arXiv},
       eprint = {2509.10346},
 primaryClass = {astro-ph.HE},
       adsurl = {https://ui.adsabs.harvard.edu/abs/2025ApJ...992...30I}
}

@BOOK{2008chcp.book.....L,
       author = {{Lide}, David R.},
        title = "{CRC Handbook of chemistry and physics: a ready-reference book of chemical and physical data}",
         year = 2008,
       adsurl = {https://ui.adsabs.harvard.edu/abs/2008chcp.book.....L}
}

@ARTICLE{2025MNRAS.543.2791Y,
       author = {{Yu}, Huan and {Fang}, Jun},
        title = "{Particle acceleration along magnetic fields as the origin of ear-like structures in supernova remnants}",
      journal = {\mnras},
         year = 2025,
        month = nov,
       volume = {543},
       number = {3},
        pages = {2791-2795},
          doi = {10.1093/mnras/staf1654},
archivePrefix = {arXiv},
       eprint = {2509.23547},
 primaryClass = {astro-ph.HE},
       adsurl = {https://ui.adsabs.harvard.edu/abs/2025MNRAS.543.2791Y}
}

@ARTICLE{1988ApJ...334..252C,
       author = {{Cioffi}, Denis F. and {McKee}, Christopher F. and {Bertschinger}, Edmund},
        title = "{Dynamics of Radiative Supernova Remnants}",
      journal = {\apj},
         year = 1988,
        month = nov,
       volume = {334},
        pages = {252},
          doi = {10.1086/166834},
       adsurl = {https://ui.adsabs.harvard.edu/abs/1988ApJ...334..252C}
}

@ARTICLE{2025ApJ...982...60P,
       author = {{Prust}, Logan J. and {Kumar}, Gabriel and {Bildsten}, Lars},
        title = "{Ejecta Wakes from Companion Interaction in Type Ia Supernova Remnants}",
      journal = {\apj},
         year = 2025,
        month = mar,
       volume = {982},
       number = {1},
          eid = {60},
        pages = {60},
          doi = {10.3847/1538-4357/adb7db},
archivePrefix = {arXiv},
       eprint = {2412.18226},
 primaryClass = {astro-ph.HE},
       adsurl = {https://ui.adsabs.harvard.edu/abs/2025ApJ...982...60P}
}

@ARTICLE{2025A&ARv..33....1R,
       author = {{Ruiter}, Ashley Jade and {Seitenzahl}, Ivo Rolf},
        title = "{Type Ia supernova progenitors: a contemporary view of a long-standing puzzle}",
      journal = {\aapr},
         year = 2025,
        month = dec,
       volume = {33},
       number = {1},
          eid = {1},
        pages = {1},
          doi = {10.1007/s00159-024-00158-9},
archivePrefix = {arXiv},
       eprint = {2412.01766},
 primaryClass = {astro-ph.SR},
       adsurl = {https://ui.adsabs.harvard.edu/abs/2025A&ARv..33....1R}
}

@ARTICLE{2024ApJ...972..200B,
       author = {{Boos}, Samuel J. and {Townsley}, Dean M. and {Shen}, Ken J.},
        title = "{Type Ia Supernovae Can Arise from the Detonations of Both Stars in a Double Degenerate Binary}",
      journal = {\apj},
         year = 2024,
        month = sep,
       volume = {972},
       number = {2},
          eid = {200},
        pages = {200},
          doi = {10.3847/1538-4357/ad5da2},
archivePrefix = {arXiv},
       eprint = {2401.08011},
 primaryClass = {astro-ph.HE},
       adsurl = {https://ui.adsabs.harvard.edu/abs/2024ApJ...972..200B}
}

@dataset{boos_2024_10515767,
  author       = {Boos, Samuel},
  title        = {dataset for "Type Ia Supernovae Can Arise from the
                   Detonations of Both Stars in a Double Degenerate
                   Binary"
                  },
  month        = jan,
  year         = 2024,
  publisher    = {Zenodo},
  doi          = {10.5281/zenodo.10515767},
  url          = {https://doi.org/10.5281/zenodo.10515767},
}

@ARTICLE{2023ApJ...950L..10S,
       author = {{Shields}, Joshua V. and {Arunachalam}, Prasiddha and {Kerzendorf}, Wolfgang and {Hughes}, John P. and {Biriouk}, Sofia and {Monk}, Hayden and {Buchner}, Johannes},
        title = "{No Surviving SN Ia Companion in SNR 0509-67.5: Stellar Population Characterization and Comparison to Models}",
      journal = {\apjl},
         year = 2023,
        month = jun,
       volume = {950},
       number = {2},
          eid = {L10},
        pages = {L10},
          doi = {10.3847/2041-8213/acd6a0},
archivePrefix = {arXiv},
       eprint = {2305.03750},
 primaryClass = {astro-ph.SR},
       adsurl = {https://ui.adsabs.harvard.edu/abs/2023ApJ...950L..10S}
}

@ARTICLE{2023OJAp....6E..28E,
       author = {{El-Badry}, Kareem and {Shen}, Ken J. and {Chandra}, Vedant and {Bauer}, Evan B. and {Fuller}, Jim and {Strader}, Jay and {Chomiuk}, Laura and {Naidu}, Rohan P. and {Caiazzo}, Ilaria and {Rodriguez}, Antonio C. and {Nagarajan}, Pranav and {Yamaguchi}, Natsuko and {Vanderbosch}, Zachary P. and {Roulston}, Benjamin R. and {G{\"a}nsicke}, Boris and {Han}, Jiwon Jesse and {Burdge}, Kevin B. and {Filippenko}, Alexei V. and {Brink}, Thomas G. and {Zheng}, WeiKang},
        title = "{The fastest stars in the Galaxy}",
      journal = {The Open Journal of Astrophysics},
         year = 2023,
        month = jul,
       volume = {6},
          eid = {28},
        pages = {28},
          doi = {10.21105/astro.2306.03914},
archivePrefix = {arXiv},
       eprint = {2306.03914},
 primaryClass = {astro-ph.SR},
       adsurl = {https://ui.adsabs.harvard.edu/abs/2023OJAp....6E..28E}
}

@ARTICLE{2024ApJ...973...65W,
       author = {{Wong}, Tin Long Sunny and {White}, Christopher J. and {Bildsten}, Lars},
        title = "{Shocking and Mass Loss of Compact Donor Stars in Type Ia Supernovae}",
      journal = {\apj},
         year = 2024,
        month = sep,
       volume = {973},
       number = {1},
          eid = {65},
        pages = {65},
          doi = {10.3847/1538-4357/ad6a11},
archivePrefix = {arXiv},
       eprint = {2408.00125},
 primaryClass = {astro-ph.SR},
       adsurl = {https://ui.adsabs.harvard.edu/abs/2024ApJ...973...65W}
}

@ARTICLE{2019ApJ...887...68B,
       author = {{Bauer}, Evan B. and {White}, Christopher J. and {Bildsten}, Lars},
        title = "{Remnants of Subdwarf Helium Donor Stars Ejected from Close Binaries with Thermonuclear Supernovae}",
      journal = {\apj},
         year = 2019,
        month = dec,
       volume = {887},
       number = {1},
          eid = {68},
        pages = {68},
          doi = {10.3847/1538-4357/ab4ea4},
archivePrefix = {arXiv},
       eprint = {1906.08941},
 primaryClass = {astro-ph.SR},
       adsurl = {https://ui.adsabs.harvard.edu/abs/2019ApJ...887...68B}
}

@Article{Hunter:2007,
  Author    = {Hunter, J. D.},
  Title     = {Matplotlib: A 2D graphics environment},
  Journal   = {Computing in Science \& Engineering},
  Volume    = {9},
  Number    = {3},
  Pages     = {90--95},
  publisher = {IEEE COMPUTER SOC},
  doi       = {10.1109/MCSE.2007.55},
  year      = 2007
}

@ARTICLE{2020SciPy-NMeth,
  author  = {Virtanen, Pauli and Gommers, Ralf and Oliphant, Travis E. and
            Haberland, Matt and Reddy, Tyler and Cournapeau, David and
            Burovski, Evgeni and Peterson, Pearu and Weckesser, Warren and
            Bright, Jonathan and {van der Walt}, St{\'e}fan J. and
            Brett, Matthew and Wilson, Joshua and Millman, K. Jarrod and
            Mayorov, Nikolay and Nelson, Andrew R. J. and Jones, Eric and
            Kern, Robert and Larson, Eric and Carey, C J and
            Polat, {\.I}lhan and Feng, Yu and Moore, Eric W. and
            {VanderPlas}, Jake and Laxalde, Denis and Perktold, Josef and
            Cimrman, Robert and Henriksen, Ian and Quintero, E. A. and
            Harris, Charles R. and Archibald, Anne M. and
            Ribeiro, Ant{\^o}nio H. and Pedregosa, Fabian and
            {van Mulbregt}, Paul and {SciPy 1.0 Contributors}},
  title   = {{{SciPy} 1.0: Fundamental Algorithms for Scientific
            Computing in Python}},
  journal = {Nature Methods},
  year    = {2020},
  volume  = {17},
  pages   = {261--272},
  adsurl  = {https://rdcu.be/b08Wh},
  doi     = {10.1038/s41592-019-0686-2},
}

@ARTICLE{2022ApJ...938..121A,
       author = {{Arunachalam}, Prasiddha and {Hughes}, John P. and {Hovey}, Luke and {Eriksen}, Kristoffer},
        title = "{A Hydro-based MCMC Analysis of SNR 0509-67.5: Revealing the Explosion Properties from Fluid Discontinuities Alone}",
      journal = {\apj},
         year = 2022,
        month = oct,
       volume = {938},
       number = {2},
          eid = {121},
        pages = {121},
          doi = {10.3847/1538-4357/ac927c},
archivePrefix = {arXiv},
       eprint = {2208.07693},
 primaryClass = {astro-ph.HE},
       adsurl = {https://ui.adsabs.harvard.edu/abs/2022ApJ...938..121A}
}

@Article{         harris2020arrayNUMPY,
 title         = {Array programming with {NumPy}},
 author        = {Charles R. Harris and K. Jarrod Millman and St{\'{e}}fan J.
                 van der Walt and Ralf Gommers and Pauli Virtanen and David
                 Cournapeau and Eric Wieser and Julian Taylor and Sebastian
                 Berg and Nathaniel J. Smith and Robert Kern and Matti Picus
                 and Stephan Hoyer and Marten H. van Kerkwijk and Matthew
                 Brett and Allan Haldane and Jaime Fern{\'{a}}ndez del
                 R{\'{i}}o and Mark Wiebe and Pearu Peterson and Pierre
                 G{\'{e}}rard-Marchant and Kevin Sheppard and Tyler Reddy and
                 Warren Weckesser and Hameer Abbasi and Christoph Gohlke and
                 Travis E. Oliphant},
 year          = {2020},
 month         = sep,
 journal       = {Nature},
 volume        = {585},
 number        = {7825},
 pages         = {357--362},
 doi           = {10.1038/s41586-020-2649-2},
 publisher     = {Springer Science and Business Media {LLC}},
 url           = {https://doi.org/10.1038/s41586-020-2649-2}
}

@ARTICLE{2018ApJ...868...90T,
       author = {{Tanikawa}, Ataru and {Nomoto}, Ken'ichi and {Nakasato}, Naohito},
        title = "{Three-dimensional Simulation of Double Detonations in the Double-degenerate Model for Type Ia Supernovae and Interaction of Ejecta with a Surviving White Dwarf Companion}",
      journal = {\apj},
         year = 2018,
        month = dec,
       volume = {868},
       number = {2},
          eid = {90},
        pages = {90},
          doi = {10.3847/1538-4357/aae9ee},
archivePrefix = {arXiv},
       eprint = {1808.01545},
 primaryClass = {astro-ph.HE},
       adsurl = {https://ui.adsabs.harvard.edu/abs/2018ApJ...868...90T}
}

@ARTICLE{2019ApJ...885..103T,
       author = {{Tanikawa}, Ataru and {Nomoto}, Ken'ichi and {Nakasato}, Naohito and {Maeda}, Keiichi},
        title = "{Double-detonation Models for Type Ia Supernovae: Trigger of Detonation in Companion White Dwarfs and Signatures of Companions{\textquoteright} Stripped-off Materials}",
      journal = {\apj},
         year = 2019,
        month = nov,
       volume = {885},
       number = {2},
          eid = {103},
        pages = {103},
          doi = {10.3847/1538-4357/ab46b6},
archivePrefix = {arXiv},
       eprint = {1909.09770},
 primaryClass = {astro-ph.HE},
       adsurl = {https://ui.adsabs.harvard.edu/abs/2019ApJ...885..103T}
}

@ARTICLE{2022ApJ...930...92F,
       author = {{Ferrand}, Gilles and {Tanikawa}, Ataru and {Warren}, Donald C. and {Nagataki}, Shigehiro and {Safi-Harb}, Samar and {Decourchelle}, Anne},
        title = "{The Double Detonation of a Double-degenerate System, from Type Ia Supernova Explosion to its Supernova Remnant}",
      journal = {\apj},
         year = 2022,
        month = may,
       volume = {930},
       number = {1},
          eid = {92},
        pages = {92},
          doi = {10.3847/1538-4357/ac5c58},
archivePrefix = {arXiv},
       eprint = {2202.04268},
 primaryClass = {astro-ph.HE},
       adsurl = {https://ui.adsabs.harvard.edu/abs/2022ApJ...930...92F}
}

@ARTICLE{2025arXiv251018800F,
       author = {{Ferrand}, Gilles and {Pakmor}, R{\"u}diger and {Fujimaru}, Yusei and {Lee}, Shiu-Hang and {Safi-Harb}, Samar and {Nagataki}, Shigehiro and {Roepke}, Friedrich K. and {Decourchelle}, Anne and {Seitenzahl}, Ivo R. and {Patnaude}, Daniel},
        title = "{The role of the secondary white dwarf in a double-degenerate double-detonation explosion, in the supernova remnant phase}",
      journal = {arXiv e-prints},
         year = 2025,
        month = oct,
          eid = {arXiv:2510.18800},
        pages = {arXiv:2510.18800},
          doi = {10.48550/arXiv.2510.18800},
archivePrefix = {arXiv},
       eprint = {2510.18800},
 primaryClass = {astro-ph.HE},
       adsurl = {https://ui.adsabs.harvard.edu/abs/2025arXiv251018800F}
}

@ARTICLE{2012A&ARv..20...49V,
       author = {{Vink}, Jacco},
        title = "{Supernova remnants: the X-ray perspective}",
      journal = {\aapr},
         year = 2012,
        month = dec,
       volume = {20},
          eid = {49},
        pages = {49},
          doi = {10.1007/s00159-011-0049-1},
archivePrefix = {arXiv},
       eprint = {1112.0576},
 primaryClass = {astro-ph.HE},
       adsurl = {https://ui.adsabs.harvard.edu/abs/2012A&ARv..20...49V}
}

@ARTICLE{2016ApJ...833...62G,
       author = {{Gray}, William J. and {Raskin}, Cody and {Owen}, J. Michael},
        title = "{Shadows of our Former Companions: How the Single-degenerate Binary Type Ia Supernova Scenario Affects Remnants}",
      journal = {\apj},
         year = 2016,
        month = dec,
       volume = {833},
       number = {1},
          eid = {62},
        pages = {62},
          doi = {10.3847/1538-4357/833/1/62},
archivePrefix = {arXiv},
       eprint = {1609.06319},
 primaryClass = {astro-ph.HE},
       adsurl = {https://ui.adsabs.harvard.edu/abs/2016ApJ...833...62G}
}

@ARTICLE{2015MNRAS.449..942P,
       author = {{Papish}, Oded and {Soker}, Noam and {Garc{\'\i}a-Berro}, Enrique and {Aznar-Sigu{\'a}n}, Gabriela},
        title = "{The response of a helium white dwarf to an exploding Type Ia supernova}",
      journal = {\mnras},
         year = 2015,
        month = may,
       volume = {449},
       number = {1},
        pages = {942-954},
          doi = {10.1093/mnras/stv337},
archivePrefix = {arXiv},
       eprint = {1410.1153},
 primaryClass = {astro-ph.SR},
       adsurl = {https://ui.adsabs.harvard.edu/abs/2015MNRAS.449..942P}
}

@ARTICLE{2024arXiv240814301X,
       author = {{XRISM Collaboration}},
        title = "{The XRISM First Light Observation: Velocity Structure and Thermal Properties of the Supernova Remnant N132D}",
      journal = {arXiv e-prints},
         year = 2024,
        month = aug,
          eid = {arXiv:2408.14301},
        pages = {arXiv:2408.14301},
          doi = {10.48550/arXiv.2408.14301},
archivePrefix = {arXiv},
       eprint = {2408.14301},
 primaryClass = {astro-ph.HE},
       adsurl = {https://ui.adsabs.harvard.edu/abs/2024arXiv240814301X}
}

@ARTICLE{2012ApJ...745...75G,
       author = {{Garc{\'\i}a-Senz}, D. and {Badenes}, C. and {Serichol}, N.},
        title = "{Is There a Hidden Hole in Type Ia Supernova Remnants?}",
      journal = {\apj},
         year = 2012,
        month = jan,
       volume = {745},
       number = {1},
          eid = {75},
        pages = {75},
          doi = {10.1088/0004-637X/745/1/75},
archivePrefix = {arXiv},
       eprint = {1110.4267},
 primaryClass = {astro-ph.SR},
       adsurl = {https://ui.adsabs.harvard.edu/abs/2012ApJ...745...75G}
}

@software{soham_mandal_2025_15595613,
  author       = {Soham Mandal and
                  Prust, Logan},
  title        = {ljprust/Sprout: v1.2-upwind},
  month        = jun,
  year         = 2025,
  publisher    = {Zenodo},
  version      = {v1.2-upwind},
  doi          = {10.5281/zenodo.15595613},
  url          = {https://doi.org/10.5281/zenodo.15595613},
  swhid        = {swh:1:dir:16ca5ce381955389c650b1dc2d678c3cca9711d0
                   ;origin=https://doi.org/10.5281/zenodo.14768191;vi
                   sit=swh:1:snp:3cb8e10beaa1ef183c2fe76ad57f0ded87ed
                   436e;anchor=swh:1:rel:26fe24c2d4227bbdd27b20bdd7a1
                   a970e7839bb0;path=ljprust-Sprout-225abd4
                  }
}

@ARTICLE{2022MNRAS.517.5260P,
       author = {{Pakmor}, R. and {Callan}, F.~P. and {Collins}, C.~E. and {de Mink}, S.~E. and {Holas}, A. and {Kerzendorf}, W.~E. and {Kromer}, M. and {Neunteufel}, P.~G. and {O'Brien}, John T. and {R{\"o}pke}, F.~K. and {Ruiter}, A.~J. and {Seitenzahl}, I.~R. and {Shingles}, Luke J. and {Sim}, S.~A. and {Taubenberger}, S.},
        title = "{On the fate of the secondary white dwarf in double-degenerate double-detonation Type Ia supernovae}",
      journal = {\mnras},
         year = 2022,
        month = dec,
       volume = {517},
       number = {4},
        pages = {5260-5271},
          doi = {10.1093/mnras/stac3107},
archivePrefix = {arXiv},
       eprint = {2203.14990},
 primaryClass = {astro-ph.SR},
       adsurl = {https://ui.adsabs.harvard.edu/abs/2022MNRAS.517.5260P}
}

@ARTICLE{2013ApJ...774...99K,
       author = {{Kerzendorf}, Wolfgang E. and {Yong}, David and {Schmidt}, Brian P. and {Simon}, Joshua D. and {Jeffery}, C. Simon and {Anderson}, Jay and {Podsiadlowski}, Philipp and {Gal-Yam}, Avishay and {Silverman}, Jeffrey M. and {Filippenko}, Alexei V. and {Nomoto}, Ken'ichi and {Murphy}, Simon J. and {Bessell}, Michael S. and {Venn}, Kim A. and {Foley}, Ryan J.},
        title = "{A High-resolution Spectroscopic Search for the Remaining Donor for Tycho's Supernova}",
      journal = {\apj},
         year = 2013,
        month = sep,
       volume = {774},
       number = {2},
          eid = {99},
        pages = {99},
          doi = {10.1088/0004-637X/774/2/99},
archivePrefix = {arXiv},
       eprint = {1210.2713},
 primaryClass = {astro-ph.SR},
       adsurl = {https://ui.adsabs.harvard.edu/abs/2013ApJ...774...99K}
}

@ARTICLE{2004Natur.431.1069R,
       author = {{Ruiz-Lapuente}, Pilar and {Comeron}, Fernando and {M{\'e}ndez}, Javier and {Canal}, Ramon and {Smartt}, Stephen J. and {Filippenko}, Alexei V. and {Kurucz}, Robert L. and {Chornock}, Ryan and {Foley}, Ryan J. and {Stanishev}, Vallery and {Ibata}, Rodrigo},
        title = "{The binary progenitor of Tycho Brahe's 1572 supernova}",
      journal = {\nat},
         year = 2004,
        month = oct,
       volume = {431},
       number = {7012},
        pages = {1069-1072},
          doi = {10.1038/nature03006},
archivePrefix = {arXiv},
       eprint = {astro-ph/0410673},
 primaryClass = {astro-ph},
       adsurl = {https://ui.adsabs.harvard.edu/abs/2004Natur.431.1069R}
}

@ARTICLE{2018ApJ...862..124R,
       author = {{Ruiz-Lapuente}, Pilar and {Damiani}, Francesco and {Bedin}, Luigi and {Gonz{\'a}lez Hern{\'a}ndez}, Jonay I. and {Galbany}, Llu{\'\i}s and {Pritchard}, John and {Canal}, Ramon and {M{\'e}ndez}, Javier},
        title = "{No Surviving Companion in Kepler's Supernova}",
      journal = {\apj},
         year = 2018,
        month = aug,
       volume = {862},
       number = {2},
          eid = {124},
        pages = {124},
          doi = {10.3847/1538-4357/aac9c4},
archivePrefix = {arXiv},
       eprint = {1711.00876},
 primaryClass = {astro-ph.SR},
       adsurl = {https://ui.adsabs.harvard.edu/abs/2018ApJ...862..124R}
}

@ARTICLE{2018MNRAS.479..192K,
       author = {{Kerzendorf}, W.~E. and {Strampelli}, G. and {Shen}, K.~J. and {Schwab}, J. and {Pakmor}, R. and {Do}, T. and {Buchner}, J. and {Rest}, A.},
        title = "{A search for a surviving companion in SN 1006}",
      journal = {\mnras},
         year = 2018,
        month = sep,
       volume = {479},
       number = {1},
        pages = {192-199},
          doi = {10.1093/mnras/sty1357},
archivePrefix = {arXiv},
       eprint = {1709.06566},
 primaryClass = {astro-ph.SR},
       adsurl = {https://ui.adsabs.harvard.edu/abs/2018MNRAS.479..192K}
}

@ARTICLE{2025arXiv250608081H,
       author = {{Hollands}, Mark A. and {Shen}, Ken. J. and {Raddi}, Roberto and {Gaensicke}, Boris T. and {Bauer}, Evan B. and {Rebassa-Mansergas}, Alberto},
        title = "{Spectroscopic and kinematic analyses of a warm survivor of a D6 supernova}",
      journal = {arXiv e-prints},
         year = 2025,
        month = jun,
          eid = {arXiv:2506.08081},
        pages = {arXiv:2506.08081},
archivePrefix = {arXiv},
       eprint = {2506.08081},
 primaryClass = {astro-ph.SR},
       adsurl = {https://ui.adsabs.harvard.edu/abs/2025arXiv250608081H}
}

@ARTICLE{2022ApJ...933L..31S,
       author = {{Shields}, Joshua V. and {Kerzendorf}, Wolfgang and {Hosek}, Matthew W. and {Shen}, Ken J. and {Rest}, Armin and {Do}, Tuan and {Lu}, Jessica R. and {Fullard}, Andrew G. and {Strampelli}, Giovanni and {Zenteno}, Alfredo},
        title = "{Searching for a Hypervelocity White Dwarf SN Ia Companion: A Proper-motion Survey of SN 1006}",
      journal = {\apjl},
         year = 2022,
        month = jul,
       volume = {933},
       number = {2},
          eid = {L31},
        pages = {L31},
          doi = {10.3847/2041-8213/ac7950},
archivePrefix = {arXiv},
       eprint = {2206.04095},
 primaryClass = {astro-ph.SR},
       adsurl = {https://ui.adsabs.harvard.edu/abs/2022ApJ...933L..31S}
}

@ARTICLE{2018ApJ...865...15S,
       author = {{Shen}, Ken J. and {Boubert}, Douglas and {G{\"a}nsicke}, Boris T. and {Jha}, Saurabh W. and {Andrews}, Jennifer E. and {Chomiuk}, Laura and {Foley}, Ryan J. and {Fraser}, Morgan and {Gromadzki}, Mariusz and {Guillochon}, James and {Kotze}, Marissa M. and {Maguire}, Kate and {Siebert}, Matthew R. and {Smith}, Nathan and {Strader}, Jay and {Badenes}, Carles and {Kerzendorf}, Wolfgang E. and {Koester}, Detlev and {Kromer}, Markus and {Miles}, Broxton and {Pakmor}, R{\"u}diger and {Schwab}, Josiah and {Toloza}, Odette and {Toonen}, Silvia and {Townsley}, Dean M. and {Williams}, Brian J.},
        title = "{Three Hypervelocity White Dwarfs in Gaia DR2: Evidence for Dynamically Driven Double-degenerate Double-detonation Type Ia Supernovae}",
      journal = {\apj},
         year = 2018,
        month = sep,
       volume = {865},
       number = {1},
          eid = {15},
        pages = {15},
          doi = {10.3847/1538-4357/aad55b},
archivePrefix = {arXiv},
       eprint = {1804.11163},
 primaryClass = {astro-ph.SR},
       adsurl = {https://ui.adsabs.harvard.edu/abs/2018ApJ...865...15S}
}

@ARTICLE{2014ApJ...785...61S,
       author = {{Shen}, Ken J. and {Bildsten}, Lars},
        title = "{The Ignition of Carbon Detonations via Converging Shock Waves in White Dwarfs}",
      journal = {\apj},
         year = 2014,
        month = apr,
       volume = {785},
       number = {1},
          eid = {61},
        pages = {61},
          doi = {10.1088/0004-637X/785/1/61},
archivePrefix = {arXiv},
       eprint = {1305.6925},
 primaryClass = {astro-ph.HE},
       adsurl = {https://ui.adsabs.harvard.edu/abs/2014ApJ...785...61S}
}

@ARTICLE{sprout,
       author = {{Mandal}, Soham and {Duffell}, Paul C.},
        title = "{SPROUT: A Moving-mesh Hydro Code Using a Uniformly Expanding Cartesian Grid}",
      journal = {\apjs},
         year = 2023,
        month = nov,
       volume = {269},
       number = {1},
          eid = {30},
        pages = {30},
          doi = {10.3847/1538-4365/acfc19},
archivePrefix = {arXiv},
       eprint = {2307.13785},
 primaryClass = {astro-ph.IM},
       adsurl = {https://ui.adsabs.harvard.edu/abs/2023ApJS..269...30M}
}

@inproceedings{access,
author = {Boerner, Timothy J. and Deems, Stephen and Furlani, Thomas R. and Knuth, Shelley L. and Towns, John},
title = {ACCESS: Advancing Innovation: NSF’s Advanced Cyberinfrastructure Coordination Ecosystem: Services \& Support},
year = {2023},
isbn = {9781450399852},
publisher = {Association for Computing Machinery},
address = {New York, NY, USA},
url = {https://doi.org/10.1145/3569951.3597559},
doi = {10.1145/3569951.3597559},
booktitle = {Practice and Experience in Advanced Research Computing},
pages = {173–176},
numpages = {4},
location = {Portland, OR, USA},
series = {PEARC '23}
}

@ARTICLE{2019PhRvL.123d1101S,
       author = {{Seitenzahl}, I.~R. and {Ghavamian}, P. and {Laming}, J.~M. and {Vogt}, F.~P.~A.},
        title = "{Optical Tomography of Chemical Elements Synthesized in Type Ia Supernovae}",
      journal = {\prl},
         year = 2019,
        month = jul,
       volume = {123},
       number = {4},
          eid = {041101},
        pages = {041101},
          doi = {10.1103/PhysRevLett.123.041101},
archivePrefix = {arXiv},
       eprint = {1906.05972},
 primaryClass = {astro-ph.SR},
       adsurl = {https://ui.adsabs.harvard.edu/abs/2019PhRvL.123d1101S}
}

@ARTICLE{2025NatAs.tmp..135D,
       author = {{Das}, Priyam and {Seitenzahl}, Ivo R. and {Ruiter}, Ashley J. and {R{\"o}pke}, Friedrich K. and {Pakmor}, R{\"u}diger and {Vogt}, Fr{\'e}d{\'e}ric P.~A. and {Collins}, Christine E. and {Ghavamian}, Parviz and {Sim}, Stuart A. and {Williams}, Brian J. and {Taubenberger}, Stefan and {Laming}, J. Martin and {Suherli}, Janette and {Sutherland}, Ralph and {Rodr{\'\i}guez-Segovia}, Nicol{\'a}s},
        title = "{Calcium in a supernova remnant as a fingerprint of a sub-Chandrasekhar-mass explosion}",
      journal = {Nature Astronomy},
         year = 2025,
        month = jul,
          doi = {10.1038/s41550-025-02589-5},
       adsurl = {https://ui.adsabs.harvard.edu/abs/2025NatAs.tmp..135D}
}

@ARTICLE{2018ApJ...860...19G,
       author = {{Garc{\'\i}a-Segura}, Guillermo and {Ricker}, Paul M. and {Taam}, Ronald E.},
        title = "{Common Envelope Shaping of Planetary Nebulae}",
      journal = {\apj},
         year = 2018,
        month = jun,
       volume = {860},
       number = {1},
          eid = {19},
        pages = {19},
          doi = {10.3847/1538-4357/aac08c},
archivePrefix = {arXiv},
       eprint = {1804.09309},
 primaryClass = {astro-ph.SR},
       adsurl = {https://ui.adsabs.harvard.edu/abs/2018ApJ...860...19G}
}

@ARTICLE{2023ApJ...949...50R,
       author = {{Raymond}, John C. and {Ghavamian}, Parviz and {Bohdan}, Artem and {Ryu}, Dongsu and {Niemiec}, Jacek and {Sironi}, Lorenzo and {Tran}, Aaron and {Amato}, Elena and {Hoshino}, Masahiro and {Pohl}, Martin and {Amano}, Takanobu and {Fiuza}, Frederico},
        title = "{Electron-Ion Temperature Ratio in Astrophysical Shocks}",
      journal = {\apj},
         year = 2023,
        month = jun,
       volume = {949},
       number = {2},
          eid = {50},
        pages = {50},
          doi = {10.3847/1538-4357/acc528},
archivePrefix = {arXiv},
       eprint = {2303.08849},
 primaryClass = {astro-ph.GA},
       adsurl = {https://ui.adsabs.harvard.edu/abs/2023ApJ...949...50R}
}

@ARTICLE{2011ApJ...732..114L,
       author = {{Lopez}, Laura A. and {Ramirez-Ruiz}, Enrico and {Huppenkothen}, Daniela and {Badenes}, Carles and {Pooley}, David A.},
        title = "{Using the X-ray Morphology of Young Supernova Remnants to Constrain Explosion Type, Ejecta Distribution, and Chemical Mixing}",
      journal = {\apj},
         year = 2011,
        month = may,
       volume = {732},
       number = {2},
          eid = {114},
        pages = {114},
          doi = {10.1088/0004-637X/732/2/114},
archivePrefix = {arXiv},
       eprint = {1011.0731},
 primaryClass = {astro-ph.HE},
       adsurl = {https://ui.adsabs.harvard.edu/abs/2011ApJ...732..114L}
}

@ARTICLE{1998JCoPh.145..511S,
       author = {{Sanders}, Richard and {Morano}, Eric and {Druguet}, Marie-Claude},
        title = "{Multidimensional Dissipation for Upwind Schemes: Stability and Applications to Gas Dynamics}",
      journal = {Journal of Computational Physics},
         year = 1998,
        month = sep,
       volume = {145},
       number = {2},
        pages = {511-537},
          doi = {10.1006/jcph.1998.6047},
       adsurl = {https://ui.adsabs.harvard.edu/abs/1998JCoPh.145..511S}
}

@ARTICLE{2025arXiv250513671L,
       author = {{Leahy}, Denis A. and {Ranasinghe}, S. and {Hansen}, J. and {Filipovi{\'c}}, M.~D. and {Smeaton}, Z.},
        title = "{Multipole Analysis and Application to Supernova Remnant X-ray Images}",
      journal = {arXiv e-prints},
         year = 2025,
        month = may,
          eid = {arXiv:2505.13671},
        pages = {arXiv:2505.13671},
archivePrefix = {arXiv},
       eprint = {2505.13671},
 primaryClass = {astro-ph.HE},
       adsurl = {https://ui.adsabs.harvard.edu/abs/2025arXiv250513671L}
}

@ARTICLE{1982ApJ...261..473T,
       author = {{Tuohy}, I.~R. and {Dopita}, M.~A. and {Mathewson}, D.~S. and {Long}, K.~S. and {Helfand}, D.~J.},
        title = "{Optical identification of Balmer-dominated supernova remnants in the Large Magellanic Cloud.}",
      journal = {\apj},
         year = 1982,
        month = oct,
       volume = {261},
        pages = {473-484},
          doi = {10.1086/160358},
       adsurl = {https://ui.adsabs.harvard.edu/abs/1982ApJ...261..473T}
}

@ARTICLE{2008ApJ...689.1089V,
       author = {{van Adelsberg}, Matthew and {Heng}, Kevin and {McCray}, Richard and {Raymond}, John C.},
        title = "{Spatial Structure and Collisionless Electron Heating in Balmer-dominated Shocks}",
      journal = {\apj},
         year = 2008,
        month = dec,
       volume = {689},
       number = {2},
        pages = {1089-1104},
          doi = {10.1086/592680},
archivePrefix = {arXiv},
       eprint = {0803.2521},
 primaryClass = {astro-ph},
       adsurl = {https://ui.adsabs.harvard.edu/abs/2008ApJ...689.1089V}
}

@ARTICLE{2013A&A...558A..25M,
       author = {{Morlino}, G. and {Blasi}, P. and {Bandiera}, R. and {Amato}, E.},
        title = "{Broad Balmer line emission and cosmic ray acceleration efficiency in supernova remnant shocks}",
      journal = {\aap},
         year = 2013,
        month = oct,
       volume = {558},
          eid = {A25},
        pages = {A25},
          doi = {10.1051/0004-6361/201322006},
archivePrefix = {arXiv},
       eprint = {1306.6454},
 primaryClass = {astro-ph.HE},
       adsurl = {https://ui.adsabs.harvard.edu/abs/2013A&A...558A..25M}
}

@ARTICLE{2018ApJ...862..148H,
       author = {{Hovey}, Luke and {Hughes}, John P. and {McCully}, Curtis and {Pandya}, Viraj and {Eriksen}, Kristoffer},
        title = "{Constraints on Cosmic-ray Acceleration Efficiency in Balmer Shocks of Two Young Type Ia Supernova Remnants in the Large Magellanic Cloud}",
      journal = {\apj},
         year = 2018,
        month = aug,
       volume = {862},
       number = {2},
          eid = {148},
        pages = {148},
          doi = {10.3847/1538-4357/aac94b},
archivePrefix = {arXiv},
       eprint = {1709.08273},
 primaryClass = {astro-ph.HE},
       adsurl = {https://ui.adsabs.harvard.edu/abs/2018ApJ...862..148H}
}

@ARTICLE{1994ApJS...92..527N,
       author = {{Nadyozhin}, D.~K.},
        title = "{The Properties of NI CO Fe Decay}",
      journal = {\apjs},
         year = 1994,
        month = jun,
       volume = {92},
        pages = {527},
          doi = {10.1086/192008},
       adsurl = {https://ui.adsabs.harvard.edu/abs/1994ApJS...92..527N}
}

@ARTICLE{2025arXiv250719722K,
       author = {{Kumar}, Gabriel and {Prust}, Logan J. and {Bildsten}, Lars},
        title = "{The First Day of a Type Ia Supernova from a Double-Degenerate Binary}",
      journal = {arXiv e-prints},
         year = 2025,
        month = jul,
          eid = {arXiv:2507.19722},
        pages = {arXiv:2507.19722},
          doi = {10.48550/arXiv.2507.19722},
archivePrefix = {arXiv},
       eprint = {2507.19722},
 primaryClass = {astro-ph.HE},
       adsurl = {https://ui.adsabs.harvard.edu/abs/2025arXiv250719722K}
}

@ARTICLE{2021ApJ...915...42K,
       author = {{Kasuga}, Tomoaki and {Vink}, Jacco and {Katsuda}, Satoru and {Uchida}, Hiroyuki and {Bamba}, Aya and {Sato}, Toshiki and {Hughes}, John P.},
        title = "{Spatially Resolved RGS Analysis of Kepler's Supernova Remnant}",
      journal = {\apj},
         year = 2021,
        month = jul,
       volume = {915},
       number = {1},
          eid = {42},
        pages = {42},
          doi = {10.3847/1538-4357/abff4f},
archivePrefix = {arXiv},
       eprint = {2105.04235},
 primaryClass = {astro-ph.HE},
       adsurl = {https://ui.adsabs.harvard.edu/abs/2021ApJ...915...42K}
}

@ARTICLE{2021ApJ...916...41S,
       author = {{Sun}, Lei and {Vink}, Jacco and {Chen}, Yang and {Zhou}, Ping and {Prokhorov}, Dmitry and {P{\"u}hlhofer}, Gerd and {Malyshev}, Denys},
        title = "{The Post-impact Evolution of the X-Ray-emitting Gas in SNR 1987A as Viewed by XMM-Newton}",
      journal = {\apj},
         year = 2021,
        month = jul,
       volume = {916},
       number = {1},
          eid = {41},
        pages = {41},
          doi = {10.3847/1538-4357/ac033d},
archivePrefix = {arXiv},
       eprint = {2103.03844},
 primaryClass = {astro-ph.HE},
       adsurl = {https://ui.adsabs.harvard.edu/abs/2021ApJ...916...41S}
}

@ARTICLE{1982ApJ...258...11H,
       author = {{Henry}, R.~B.~C. and {MacAlpine}, G.~M.},
        title = "{The Crab Nebula. II - A photoionization model analysis for the filaments}",
      journal = {\apj},
         year = 1982,
        month = jul,
       volume = {258},
        pages = {11-21},
          doi = {10.1086/160044},
       adsurl = {https://ui.adsabs.harvard.edu/abs/1982ApJ...258...11H}
}

@ARTICLE{2008Sci...320.1195K,
       author = {{Krause}, Oliver and {Birkmann}, Stephan M. and {Usuda}, Tomonori and {Hattori}, Takashi and {Goto}, Miwa and {Rieke}, George H. and {Misselt}, Karl A.},
        title = "{The Cassiopeia A Supernova Was of Type IIb}",
      journal = {Science},
         year = 2008,
        month = may,
       volume = {320},
       number = {5880},
        pages = {1195},
          doi = {10.1126/science.1155788},
archivePrefix = {arXiv},
       eprint = {0805.4557},
 primaryClass = {astro-ph},
       adsurl = {https://ui.adsabs.harvard.edu/abs/2008Sci...320.1195K}
}

@ARTICLE{2021ApJ...923..141L,
       author = {{Li}, Chuan-Jui and {Chu}, You-Hua and {Raymond}, John C. and {Leibundgut}, Bruno and {Seitenzahl}, Ivo R. and {Morlino}, Giovanni},
        title = "{Forbidden Line Emission from Type Ia Supernova Remnants Containing Balmer-dominated Shells}",
      journal = {\apj},
         year = 2021,
        month = dec,
       volume = {923},
       number = {2},
          eid = {141},
        pages = {141},
          doi = {10.3847/1538-4357/ac2c04},
archivePrefix = {arXiv},
       eprint = {2110.09250},
 primaryClass = {astro-ph.HE},
       adsurl = {https://ui.adsabs.harvard.edu/abs/2021ApJ...923..141L}
}

@ARTICLE{2022ApJ...935...78W,
       author = {{Williams}, Brian J. and {Ghavamian}, Parviz and {Seitenzahl}, Ivo R. and {Reynolds}, Stephen P. and {Borkowski}, Kazimierz J. and {Petre}, Robert},
        title = "{Evidence for a Dense, Inhomogeneous Circumstellar Medium in the Type Ia SNR 0519-69.0}",
      journal = {\apj},
         year = 2022,
        month = aug,
       volume = {935},
       number = {2},
          eid = {78},
        pages = {78},
          doi = {10.3847/1538-4357/ac81ca},
archivePrefix = {arXiv},
       eprint = {2207.08724},
 primaryClass = {astro-ph.GA},
       adsurl = {https://ui.adsabs.harvard.edu/abs/2022ApJ...935...78W}
}

@ARTICLE{2025arXiv250812529W,
       author = {{Wong}, Tin Long Sunny and {Bildsten}, Lars},
        title = "{Mass Loss and Subsequent Thermal Evolution of Surviving Helium White Dwarfs Shocked by Thermonuclear Supernovae}",
      journal = {arXiv e-prints},
         year = 2025,
        month = aug,
          eid = {arXiv:2508.12529},
        pages = {arXiv:2508.12529},
archivePrefix = {arXiv},
       eprint = {2508.12529},
 primaryClass = {astro-ph.SR},
       adsurl = {https://ui.adsabs.harvard.edu/abs/2025arXiv250812529W}
}
\bibliographystyle{aasjournal}


\end{document}